\documentclass[a4paper,twocolumn,11pt, accepted=2023-07-07]{quantumarticle}
\pdfoutput=1

\usepackage{tikz}

\usepackage{etoolbox}
\patchcmd{\section}
  {\centering}
  {\raggedright}
  {}
  {}
\patchcmd{\subsection}
  {\centering}
  {\raggedright}
  {}
  {}
  
\usepackage{siunitx} % SI units

\usepackage{algorithm} % HDRG pseudocode
\usepackage[noend]{algpseudocode} % HDRG pseudocode

\usepackage{tcolorbox}
\usepackage[sort&compress, numbers]{natbib}
\usepackage{soul}
\usepackage{physics}
\usepackage{graphicx}
\usepackage{epstopdf}
\usepackage{verbatim}
\usepackage{float}
\usepackage{sidecap}
\sidecaptionvpos{figure}{c}
\usepackage{lipsum}
\usepackage{capt-of}
\usepackage{setspace}
\usepackage{dcolumn}
\usepackage{bm}
\usepackage{amssymb}
\usepackage{amsmath}
\usepackage{mathtools}
\usepackage{color}
\usepackage{multirow}
\usepackage{lipsum}

\usepackage[colorlinks=true, urlcolor=blue, pdfborder={0 0 0}]{hyperref}
\hypersetup{
linkcolor=blue,
citecolor=blue
}

\begin{document}

\title{A scalable and fast artificial neural network syndrome decoder for~surface~codes} % ~ groups words

\author{Spiro Gicev} \email{gicevs@unimelb.edu.au} \affiliation{Center for Quantum Computation and Communication Technology, School of Physics, University of Melbourne, Parkville, 3010, VIC, Australia.}

\author{Lloyd C.L. Hollenberg} \email{lloydch@unimelb.edu.au} \affiliation{Center for Quantum Computation and Communication Technology, School of Physics, University of Melbourne, Parkville, 3010, VIC, Australia.}

\author{Muhammad Usman} \email{musman@unimelb.edu.au} \affiliation{Center for Quantum Computation and Communication Technology, School of Physics, University of Melbourne, Parkville, 3010, VIC, Australia.} \affiliation{School of Computing and Information Systems, Melbourne School of Engineering, University of Melbourne, Parkville, 3010, VIC, Australia}
\affiliation{Data61, CSIRO, Clayton, 3168, VIC, Australia}

\maketitle
% \onecolumngrid
\begin{abstract}
\noindent
    Surface code error correction offers a highly promising pathway to achieve scalable fault-tolerant quantum computing. When operated as stabilizer codes, surface code computations consist of a syndrome decoding step where measured stabilizer operators are used to determine appropriate corrections for errors in physical qubits. Decoding algorithms have undergone substantial development, with recent work incorporating machine learning (ML) techniques. Despite promising initial results, ML-based syndrome decoders are still limited to small scale demonstrations with low latency and are incapable of handling surface codes with boundary conditions and various shapes needed for lattice surgery and braiding. Here, we report the development of a scalable and fast syndrome decoder powered by an artificial neural network (ANN) which is capable of decoding surface codes of arbitrary shape and size with data qubits suffering from a variety of noise models including depolarising errors, biased noise, and spatially inhomogeneous noise. The decoding process involves syndrome processing by an ANN decoder followed by a mop-up step to correct any residual errors. Based on rigorous training over 50 million random quantum error instances, our ANN decoder is shown to work with code distances exceeding 1000 (more than 4 million physical qubits), which is the largest ML-based decoder demonstration to-date. The established ANN decoder demonstrates an execution time in principle independent of code distance, implying that its implementation on dedicated hardware could potentially offer surface code decoding times of O($\mu$sec), commensurate with the experimentally realisable qubit coherence times. With the anticipated scale-up of quantum processors within the next decade, their augmentation with a fast and scalable syndrome decoder such as developed in our work is expected to play a decisive role towards experimental implementation of fault-tolerant quantum information processing.
\end{abstract}

\begin{figure*}[htbp]
\begin{center}
\includegraphics[width=16cm]{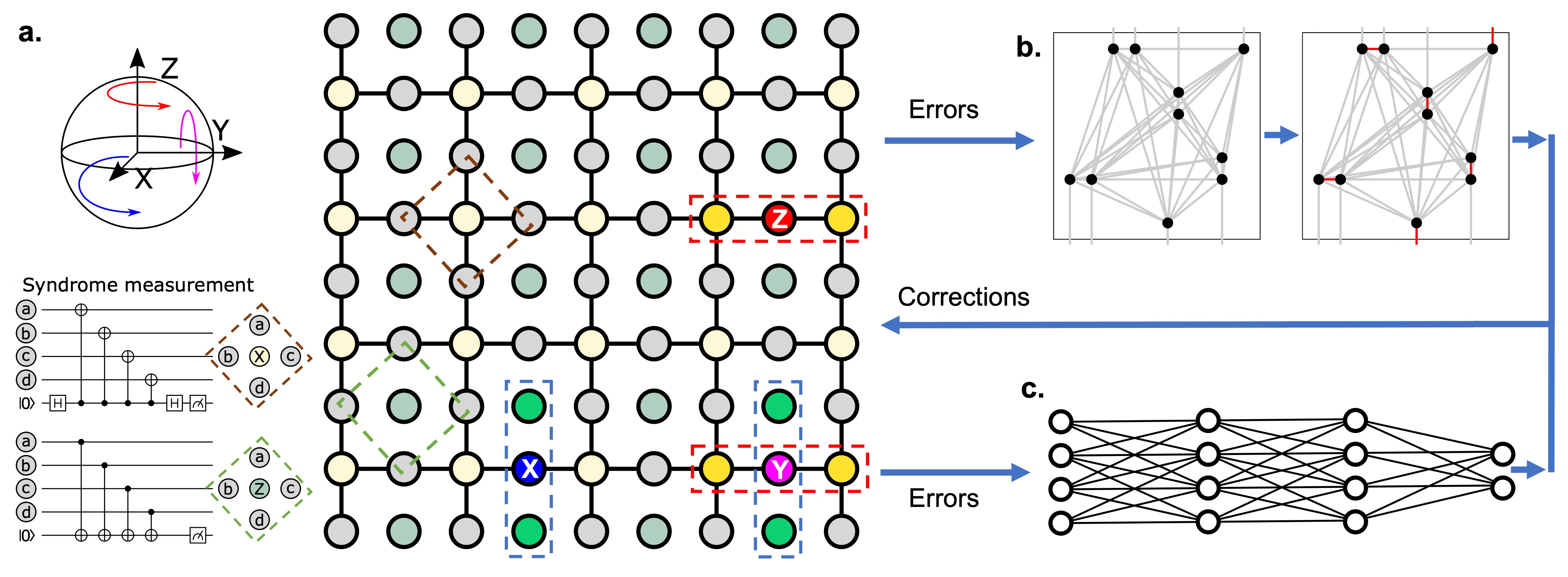}
\captionof{figure}{\textbf{Overview of surface code quantum error correction.} \textbf{a.} A schematic representation of a surface code with depolarising errors is illustrated. The top left plot shows X, Z, and Y qubit rotations on a Bloch sphere, which are examples of depolarising errors on data qubits. The bottom left circuits indicate ancillary X and Z stabilizer qubits interacting with neighbouring data qubits (labelled as a, b, c, and d) in order to measure stabilizer operators. The plot on right shows a two-dimensional array of qubits encoded in a surface code. Vertex operators (X stabilizer generators) and plaquette operators (Z stabilizer generators) are coloured in yellow and green, respectively. Bland and vibrant shades of these colours are given to +1 or \( -1 \) eigenvalue stabilizer generators, respectively. Data qubits are coloured grey, blue, magenta or red depending on whether they have had no error, an X error, a Y error or a Z error applied. This colouring convention will persist throughout this work. \textbf{b.} Visualization of a graph based decoding technique such as minimum weight perfect matching (MWPM) conventionally used in the literature, which suggests error corrections by matching pairs of \( -1 \) stabilizers. \textbf{c.} Illustration of an efficient artificial neural network based decoding scheme such as developed in this work, which predicts appropriate data qubit corrections based on forward propagation through a neural network trained by supervised learning.}
\label{fig:1}
\end{center}
\end{figure*}

\section{Introduction}
A scalable error-corrected quantum computer is anticipated to outperform classical supercomputers by efficiently performing complex computations with implications for many areas of research and development including cryptography, quantum chemistry, drug design, and optimization problems \cite{pirandola2020advances, cao2019quantum, orus2019quantum}. However, the current generation of quantum devices are inhibited from executing large circuit depth algorithms needed to demonstrate quantum advantage for practical applications due to noise, which is inherently present in all quantum systems due to factors such as fabrication imperfections, control errors and interactions with the environment. Developments in theoretical quantum error correction suggest that fault tolerant quantum computing (FTQC), capable of tackling real-world problems, is possible if enough physical qubits are available to construct error protected logical qubits, and as long as error rates are achieved beneath a particular threshold \cite{gidney2019factor, lee2020even, sanders2020compilation}. One of the most promising error correction schemes for fault tolerance is the surface code \cite{dennis2002topological}, which exploits topological properties of a qubit system, and has recently been shown as experimentally viable \cite{andersen2020naturephysics, chen2021nature}. A scalable experimental implementation of a surface code scheme will be a decisive step towards practical quantum computing. Such an implementation would benefit from both a reduction in number of physical qubits per logical qubit as well as faster operational times to enable error corrections within the limits of experimentally realisable coherence times. However, attendant to fast quantum operation times is the requirement of fast syndrome decoding to allow the quantum error correction procedure to keep up with the quantum processor itself. This is the key open problem we address in this work by exploiting the computational efficiency of a machine learning (ML) approach to establish a fast and scalable surface code error correction framework, which flexibly works with many different boundary conditions and code shapes.
\\ \\
\noindent
Figure~\ref{fig:1} (a) schematically illustrates the layout of a surface code scheme, where data qubits are surrounded by ancilla qubits. During the operation of surface codes, ancilla qubits facilitate syndrome measurements which provide information to indicate where errors may have occurred in physical qubits. The outcomes of syndrome measurements are processed by classical algorithms, known as syndrome decoders, which identify the most appropriate correction operations (such as illustrated in Figure~\ref{fig:1} (b)). A great deal of work has been done on the design of accurate surface code decoding algorithms such as minimum weight perfect matching (MWPM) and renormalisation group (RG) \cite{fowler2010surface, fowler2012towards, fowler2013optimal, watson2015fast, duclos2010fast}, as well as the development and improvement of fault tolerant surface code computational structures \cite{raussendorf2007fault,litinski2019game}. Appendix \ref{Appendix Literature Survey} provides a quick review of the decoder literature. Despite many years of research, even the best available decoder algorithms are generally slow, which limits the capability of surface codes for experimental devices. For computations at scales relevant to unambiguously demonstrate quantum advantage, efficient decoding methods need to be developed with improvements in latency, scalability and adaptability to hardware \cite{fowler2012towards, varsamopoulos2017decoding, davaasuren2020general}.
\\ \\
\noindent
An alternative approach for syndrome decoder construction is to exploit ML techniques to train a neural network to perform the decoding task as illustrated in Figure~\ref{fig:1} (c). Recent research has reported the design of artificial neural network (ANN) syndrome decoders \cite{torlai2017neural, krastanov2017deep, varsamopoulos2017decoding, baireuther2018machine, davaasuren2020general, bhoumik2021efficient, sweke2020reinforcement, matekole2022decoding, overwater2022neural}, with preliminary studies demonstrating that ANN decoders could offer highly promising performance when applied to small-scale surface codes. Many varieties of neural networks have attempted to be trained to act as efficient surface code decoders, which include dense networks \cite{varsamopoulos2017decoding}, networks with convolutional layers followed by dense layers \cite{davaasuren2020general}, restricted Boltzmann machine networks \cite{torlai2017neural}, and networks utilizing long short-term memory (LSTM) layers \cite{baireuther2018machine}. Trade-offs involving decoding latency, decoding threshold and decoding scalability have been discussed in the literature (see for example Appendix E in Ref. \cite{Meinerz2021arxiv}), with most decoders designed in the pursuit of thresholds competitive with non-ML decoders, but limited to small-scale surface code demonstrations (less than distance 100 codes).
\\ \\
\noindent
Meinerz \textit{et al.} \cite{Meinerz2021arxiv} recently reported a neural network based decoder with a divide-and-conquer approach, which was shown to work with code distances up to 255. They investigated the toric code, featuring no boundaries (limited to full periodicity) \cite{Meinerz2021arxiv}. Decoding systems with boundaries is needed for surface codes on a plane, rather than a torus. The ability to decode with a variety of different boundaries is also a requirement for the ability to implement a universal set of logical operations. Sweke \textit{et al.} \cite{sweke2020reinforcement} showed that a general reinforcement learning approach could be applied to decoding problems associated with a large class of codes and error models. They demonstrated the approach for distance 5 surface code for error models including the phenomenological noise model. Bhoumik \textit{et al.} \cite{bhoumik2021efficient} developed upon combining ``high-level" and ``low-level" approaches to decoding using neural networks \cite{varsamopoulos2019comparing}, showing that such approaches can be used in tandem. Matekole \textit{et al.} \cite{matekole2022decoding} investigated the possibility of constructing decoders which adapt to changes device noise patterns without needing to train a new model from scratch. Such changing conditions are present in real quantum devices. Overwater \textit{et al.} \cite{overwater2022neural} investigated optimization of hardware implementations of ``high-level" decoders and showed strong evidence that ``high-level" decoders could satisfy the decoding needs of surface codes with distances up to 9. While neural network decoding of surface codes continues to develop in many interesting directions, the need for decoders to be applicable to much larger distances and support multiple boundary configurations has not yet been fully addressed.
\\ \\
\noindent
Our work goes well beyond the existing literature in terms of scalability and generalization to different boundary conditions. In this work we report the development of a highly versatile ANN decoder, which is capable of outperforming conventional non-ML decoders at large code distances ($\geq$1000). We demonstrate the working of our ANN decoder for surface code implementations with distances reaching 1025 (more than four million physical qubits) and show that the established ANN decoder can process surface codes with logical qubits based on defects, braiding, and lattice surgery structures of variable shapes and sizes suffering from a variety of noise models including uniformly distributed depolarizing errors, biased noise and spatially inhomogeneous noise. Such performance was achieved without further training on particular instances of each code distance, boundary structure, and error model. 
\\ \\
\noindent
The construction of the efficient ANN decoder in our work is achieved by utilizing a fully convolutional neural network (CNN) structure which is able to correct errors accurately when given local boundary information, including systems which were not part of the training process.  A heuristic training data modification procedure was used to mitigate the negative impact of symmetries of multi-label classification during the training of the neural network. In contrast to a dense neural network generally applied in ML-based decoders, our CNN structure is based on neuron connectivity with fixed proximity which results in an execution time in principle constant with respect to code distance. This allowed syndrome decoding for surface codes consisting of multi-million physical qubits. After an exhaustive processing through ANN passes, sparse residual syndromes were decoded with the non-ML Hard Decision Renormalization Group (HDRG) \cite{watson2015fast} algorithm which serves as a final mop-up step. The implementation of our ANN+HDRG decoder with nominal compute power (without any particular focus on resource optimisation) showed that it provides a better scaling of operation time as a function of code distance when compared to an efficient and fast implementation of the MWPM algorithm \cite{Higgott2021arxiv}. It is expected that the established ANN decoder could offer significantly faster decoding times when implemented with dedicated hardware such as field programmable gate arrays (FPGAs) or application specific integrated circuits (ASICs) \cite{chamberland2018deep}. A future implementation of our ANN decoder on such dedicated hardware will offer the potential to achieve surface code decoding times of O($\mu$sec) while processing code distances exceeding 1000, achieving a crucial milestone towards experimental implementation of fault-tolerant quantum information processing.

\section{Results and Discussion}
\subsection{Surface Code Overview and Layout}
Surface codes, which are based on well-established stabilizer formalism \cite{gottesman1997stabilizer}, are capable of facilitating FTQC algorithms. After years of theoretical developments, only recently have surface code experimental demonstrations, albeit at small scale, been reported on superconducting quantum devices \cite{andersen2020naturephysics, chen2021nature}. Additionally, a number of theoretical proposals have also been reported to incorporate the surface code error correction scheme in scalable semiconductor qubit systems \cite{hill2015scienceadv, Pica_PRB_2016, hill2021arXiv} and superconducting quantum devices \cite{chamberland2020prx}, indicating potential for larger sized demonstrations in the coming years. One of the promising characteristics of surface code based error correction schemes is the associated high threshold which is 15-18\% ($\approx$0.5\%) without (with) measurement errors \cite{dennis2002topological, fowler2010surface, bombin2012prx, ashleypra2014, Wang_PRA_2011}. However, an equally important challenge is its fast implementation to ensure that syndrome processing and correction operations finish within the limitation of hardware coherence times, which is the main focus of our work while implementing the ANN component of the decoder.
\\ \\
\noindent

\begin{figure}[htp]
\begin{center}
\resizebox{80mm}{!}{\includegraphics{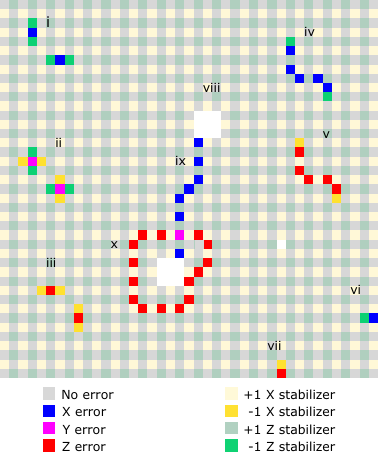}}
\end{center}
\caption{\textbf{Surface code labelling scheme.} The state of a surface code represented on a 2-dimensional grid. Each cell is coloured according to the presence of errors for data qubits and the value of stabilizer measurements for ancilla qubits. Error patterns: (i) isolated X errors, (ii) isolated Y errors, (iii) isolated Z errors, (iv) multiple adjacent X errors, (v) multiple adjacent Z errors, (vi) boundary X errors, (vii) boundary Z error, (viii) Z-cut logical qubit, (ix) double Z-cut logical qubit X operator and (x) double Z-cut logical qubit Z operator. The color coding scheme for both data and ancilla qubits is provided in the legend.
}\label{fig:2}
\end{figure}

\noindent
Figure~\ref{fig:2} illustrates the layout of the surface code compatible with a square lattice of physical qubits with controllable nearest neighbour interactions. The code space is defined by the surface code's stabilizer group. The stabilizer group (an Abelian group) can be efficiently described by its generators, which when multiplied together in all possible combinations construct the full stabilizer group. Away from edges, the stabilizer generators can be defined as the X-type ``vertex'' operators,
\begin{equation}
    V_{i,j} = X_{i-1,j} X_{i,j-1} X_{i,j+1} X_{i+1,j},
\end{equation}
\noindent
and Z-type ``plaquette'' operators,
\begin{equation}
    P_{i,j} = Z_{i-1,j} Z_{i,j-1} Z_{i,j+1} Z_{i+1,j}, 
\end{equation}
\\ \\
\noindent
with their names deriving from the earliest descriptions of surface codes \cite{dennis2002topological}. Operator subscripts denote sites of the lattice (where $i$ and $j$ refer to the row number and column number respectively). Figure~\ref{fig:2} also outlines details of our notations, where a square shaped surface code appears with Z-type stabilizer generators found at odd rows and even columns (green cells), and X-type stabilizer generators found at even rows and odd columns (yellow cells). At boundaries, stabilizer generators need to be defined differently to avoid acting on qubits which aren't present. This is done by simply removing the X or Z components of operators which act on absent qubits. Data qubits (grey cells which turn red, blue or magenta when errors occur) fill up the remainder of the board at cells with the same parity of their row and column number, forming a regular square lattice. It can be shown that the resultant set of stabilizer generators mutually commute and hence that their simultaneous measurement is permitted. The measurements of X and Z stabilizers could be performed from quantum circuits shown in Figure \ref{fig:1} (a). The collective result of all stabilizer generator measurements is known as the error syndrome.
\\ \\
\noindent
In this work, we train our decoder using the depolarizing error model where data qubits experience X, Y or Z errors with equal probability. However, the performance evaluation of the decoder was carried out based on a variety of noise configurations including depolarising, spatially inhomogeneous and biased noise which will be discussed in the Decoder Performance section. Depolarising errors cause qubits to experience X, Y, or Z errors with equal probability p/3, uniformally distributed with a fixed rate on all data qubits. Inhomogeneous noise models allow the rate of depolarisation to be a function of a qubit position in the lattice. Finally, biased noise models allow qubits to experience particular errors with higher probability, such as dominant phase or Z errors. After a Pauli string of errors has been applied, the resultant state may be written as $e_i |\psi\rangle$, where $|\psi\rangle$ is the original state and $e_i$ is a stochastic error which has occurred. The state of the data qubits of the code can be efficiently simulated, for example by keeping track of the type of errors which have occurred. The goal of the error correction scheme is to remedy errors in a way most likely to return all qubits to an error free state or in a logically equivalent state. As direct measurement of qubits in general destructively collapses their state, any errors present in the state of qubits are hidden and must be discerned by other means. Individual X, Y or Z errors on data qubits anti-commute with adjacent stabilizer operators. This causes the measured value of those stabilizer operators to be inverted. For simplicity we will assume that each stabilizer generator will always be initially measured once to return a \( +1 \) measurement result. Stabilizers with a \( +1 \) measurement result will be colour coded with the pale shades of green and yellow shown in Figure~\ref{fig:2}. Examples of individual data qubit errors can be seen in Figure \ref{fig:2} and their associated \( -1 \) stabilizer measurements, which appear as more vibrantly coloured stabilizer cells. When multiple data qubit errors occur adjacently, some stabilizer operators may have their value inverted an even number of times. In this case the only stabilizer operators which have their value changed are those at the endpoints of the error chain. Also, at boundaries individual errors can sometimes only anti-commute with a single stabilizer operator, as the lattice abruptly stops.
\\ \\
\noindent
Errors in the form of stabilizer generators or elements of the stabilizer group commute with all stabilizer generators and hence do not change any stabilizer operator measurements. This means that the overall result of errors of this type is to possibly apply a global phase to the encoded state, leaving any encoded information unaltered. Thus, errors of the form of elements of the stabilizer group can be applied liberally whenever convenient. This can be used to generate sets of homologically equivalent errors \cite{dennis2002topological}. Sets of homologically equivalent errors form the homological equivalence classes. It must be noted that homologically equivalent chains of errors always share the same boundaries (and hence change the same stabilizer operator measurements) but not all error chains with the same boundaries are homologically equivalent. 

\noindent
\\
\noindent
Lastly, some operators do exist that commute with all stabilizer generators but are not members of the stabilizer group. These operators are logical operators and change the logical state of encoded logical qubits. Under the lattice surgery scheme \cite{litinski2019game}, encoded qubits are of the boundary type, where their logical Pauli operators consist of vertical and horizontal stripes of errors. Multiple patches can be placed on a larger board of physical qubits to allow computations with multiple logical qubits. Patches may be of irregular shapes when logical qubits are kept in memory, or when participating in joint measurements \cite{gidney2019factor, Fowler_lowoverhead_2019}. Figure~\ref{fig:2} exhibits white patches which define defect based logical qubits. Under the braiding scheme \cite{raussendorf2007fault}, these defect type logical qubits are used to process logical information, where patches of qubits within the defect area temporarily cease from participating in syndrome measurements. In such a scheme, pairs of defects can be used to implement two qubit logical gates. Logical Pauli operators are then any operators which connect boundaries of the same type or which encircle one logical defect of the pair. Importantly, it can be shown that loops of operators which encircle both of the pair of defects result in logical operators which act trivially on states within the code space (equivalent to an identity operation). The goal of fault tolerant quantum computing is to run algorithms on logical qubits where non-trivial logical operators are applied to logical qubits with nearly 100\% probability when required by the quantum algorithm being executed. Equally important is the prevention of non-trivial logical operations occurring due to noise or errors. We note that this work will not discuss how to initialize, measure and perform other gates on logical qubits beyond Pauli X and Z gates, however construction of general logical gate sets have been discussed in earlier work \cite{raussendorf2007fault, litinski2019game, fowler2012surface}. It is also noted that our ANN decoder is able to process surface codes of arbitrary shape and size without explicit knowledge of such structures in the training process, which are fundamental to implement braiding and lattice surgery operations. 

\begin{figure*}[htp]
\begin{center}
\resizebox{150mm}{!}{\includegraphics{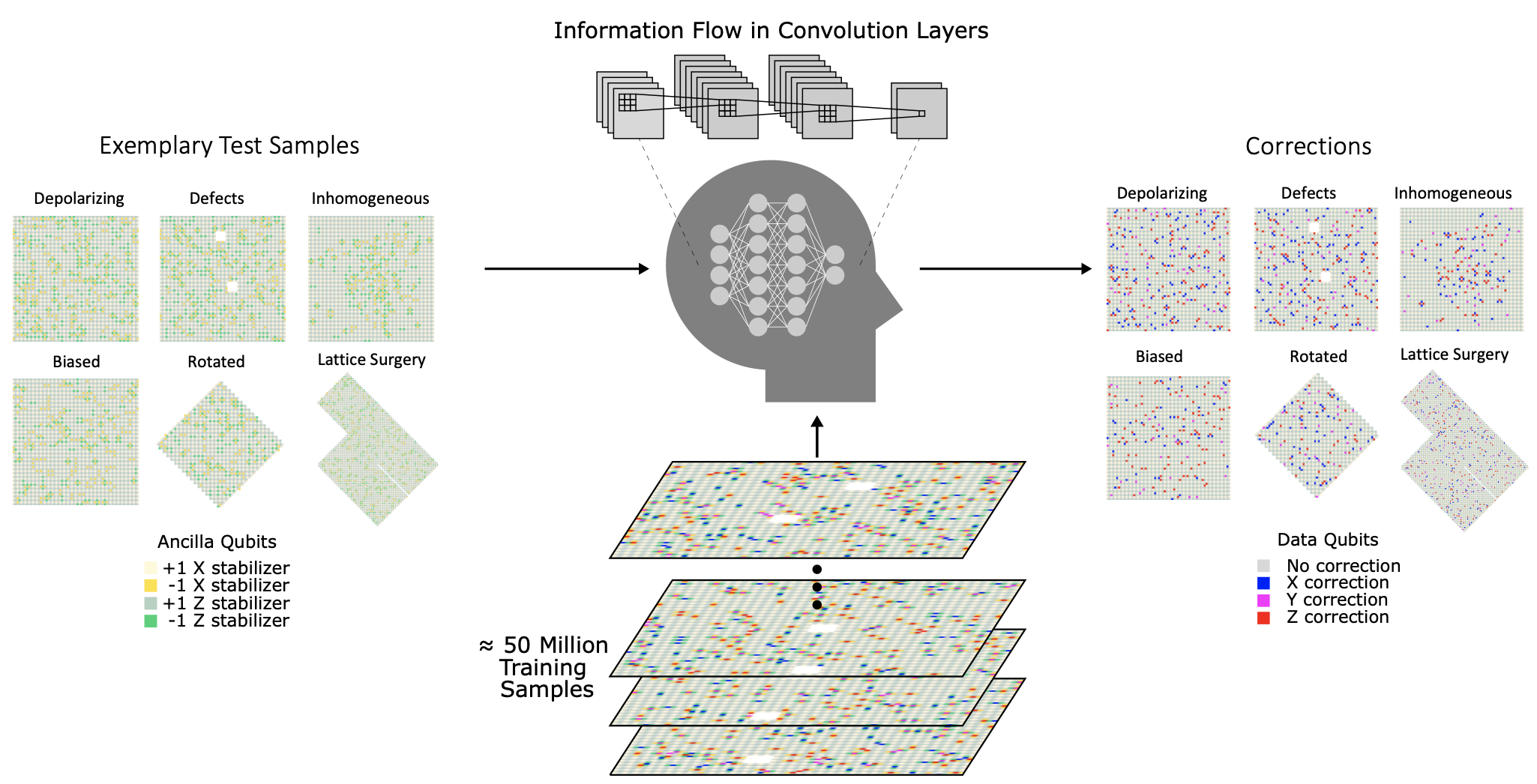}}
\end{center}
\caption{\textbf{Neural network decoding schematic.} A schematic diagram illustrating the artificial neural network decoder construction in our work. The decoder is trained through supervised learning by generating 50 million square surface code boards (training data) with random distributions of depolarising errors. The neural network consists of four layers with multi-channel input and output layers. Errors on data qubits cause non-trivial syndrome to be observed. The trained ANN was tested on a septate data set consisting of a variety of surface code boards including boards with randomly distributed homogeneous, biased and spatially inhomogeneous noise, as well as rotated surface codes and surface codes with arbitrary boundaries suitable for lattice surgery and braiding. In all cases, the ANN decoder accurately predicted appropriate correction operations on data qubits based on the syndrome measurements with low latency. 
}\label{fig:3}
\end{figure*}

\subsection{Decoder Construction}
Decoding is the process of discerning appropriate correction operations based on syndrome measurements indicative of errors in physical qubits. This is repeated multiple times throughout the process of fault tolerant quantum computation and is a major source of computational overhead compared to quantum algorithms with no error correction. The task of optimal decoding under a given error model assumption amounts to returning the appropriate correction operations most likely to return the system to the code space with no undesired logical operations being applied. In this work, we have designed a new syndrome decoder based on supervised training of an artificial neural network.
\\ \\
\noindent
Figure~\ref{fig:3} schematically illustrates the ANN decoder structure developed in our work, which is capable of processing syndrome measurements from surface codes of variable shape and size. The decoder is also able to process a variety of noise models such as uniformly distributed depolarising noise, spatially inhomogeneous noise and biased noise. To help a broader non-ML readership, a quick introduction of artificial neural networks is provided in Appendix \ref{Appendix Introduction to Artificial Neural Networks}. The construction of the ANN decoder consists entirely of locally connected neural network layers as highlighted in Appendix Figure \ref{fig:decoder_layers}. This is in contrast to previous decoders reported in the literature, which also make use of convolutional layers, but have dense final layers \cite{davaasuren2020general}.  The fully convolutional decoder implemented in our work uses a four channel input layer where the first two channels correspond to the X and Z syndrome information and the last two channels correspond to the X and Z boundary qubit information \cite{ni2020neural} (an example is shown in Appendix Figure \ref{fig:input_example}).  Boundary qubits are defined as data qubits which when acted upon by an X or Z errors cause exactly one stabilizer generator to change values. This provides sufficient information to identify appropriate correction operations based on the local syndrome environment and local boundaries of the code. The explicit inclusion of boundary information allows the decoder to process surface codes of varying shapes and sizes during the training and testing phases. These generalization properties are desirable for a decoder to process complex fault-tolerant logical operations, and have never before been demonstrated with a machine learning decoder.
\noindent
\\ \\
\noindent
The next two layers of the neural network, the hidden layers, are convolutional layers. The use of convolutional layers allows the first hidden layer to be interpreted to consist of a multi-channel image containing local information about the syndrome and boundary environment. The locality of the information is defined by the size of the convolutional kernels. Larger convolutional kernels result in information being able to be transmitted about the presence of larger clusters of errors. An alternate way to achieve the ability to identify larger clusters is to use additional hidden layers. An effect resembling a light-cone occurs where the deeper hidden layers consist of neurons which depend on a larger window of local observation of earlier layers, as shown in Figure~\ref{fig:3} and also in Appendix Figure \ref{fig:decoder_layers}. The output layer is again another convolutional layer with two kernels being trained. Each kernel results in a channel of the layer and thus the output layer will consist of two channels. These two channels are interpreted to contain the probabilities of the associated qubits requiring an X or Z correction operation. As the output will be interpreted as a probability, an activation function such as a sigmoid is used to ensure that the output is between zero and one. Finally, all the data qubits with a probability larger than 0.5 of being in error are recommended to have the appropriate correction operations applied. This is in contrast to some previous neural decoders reported in the literature, which encourage sampling from the distribution of the output until a set of errors has been found which are exactly consistent with the syndromes measured \cite{krastanov2017deep}. Such a selection process, which requires a large number of samples exponentially growing with the number of data qubits in the surface code, has not been considered in our work to enable decoding operation at code distances beyond 1000.
\\ \\
\noindent
In summary, the established ANN can be interpreted as an image to image network where the input image consists of four channels and the output image consists of two channels (see Appendix Figure \ref{fig:decoder_layers}). The output layer consisting of a sigmoid activation function leads to the behaviour of this network easily understood as qubit-wise error state logistic regression. All prior layers use ReLU activation functions. The use of exclusively convolutional layers allows the network to operate on surface code boards of arbitrary shape and size, so long as the input channels are constructed to adequately convey the required boundary information. Two hidden layers were used, each containing nine channels, and all kernels were 11$\times$11 in size (more detail regarding hyperperameter choices is given in Appendix \ref{Appendix Neural Decoder Construction}). The use of three convolutional layers results in a local receptive field with a radius of 15 cells. Using kernels of larger linear dimension or using more hidden layers would each linearly increase the local receptive field radius. To have a receptive field large enough to fully decode a surface code of distance d requires a number of hidden layers which is O(d) and hence such a decoder would have a time complexity of O(d). Such a decoder would develop behaviour resembling that of a cellular automaton. However, it stands to be shown whether or not such constructions are possible, with a possible issue being memory requirements scaling quadratically with distance. This is left for future work. The construction of the decoder in our work is based on concatenation of a purely convolutional neural network with a global decoder to correct any residual errors (see Appendix \ref{Appendix Neural Decoder Construction} for further details and a flow chart diagram is plotted in Appendix Figure \ref{fig:just_flowchart}). 

A global decoder acting as a ``mop-up'' decoder is required due to the finite receptive field when using exclusively convolutional layers. For any non-zero error rate there exists a possibility that an error chain will occur which is larger than the receptive field of the ANN decoder and hence will be unable to be corrected. This can also occur when -1 stabilizer patterns occur for which the ANN could not develop a preferred decoding strategy, which can in particular occur for some highly symmetric syndromes. One way of resolving this issue is to apply a final step in the ANN decoding process where a global decoder (capable of giving corrections to any syndrome due to full system information) applies any final required corrections. This task was given the name ``mop-up'' decoding, and is distinct from ordinary decoding in that the syndromes given often correspond to highly correlated residual errors, however sparse and in low quantities. In this regime, different decoding strategies may be close to optimal, as they do not necessarily need to optimized to be able to effectively decode uniform depolarizing noise. In this work, a HDRG decoder \cite{watson2015fast} was chosen for this task. HDRG decoders calculate corrections by iteratievly clustering syndromes into bigger groups, until individual groups can be corrected individually by either containing an even number of -1 stabilizers or being sufficiently close to a boundary. More details are given in Appendix \ref{Appendix HDRG Decoder}. Due to its local and iterative behaviour, the HDRG decoder was chosen as a good candidate for applying quick final corrections on surface codes with well separated correlated chains of errors.

\begin{figure}[htbp]
\begin{center}
\resizebox{80mm}{!}{\includegraphics{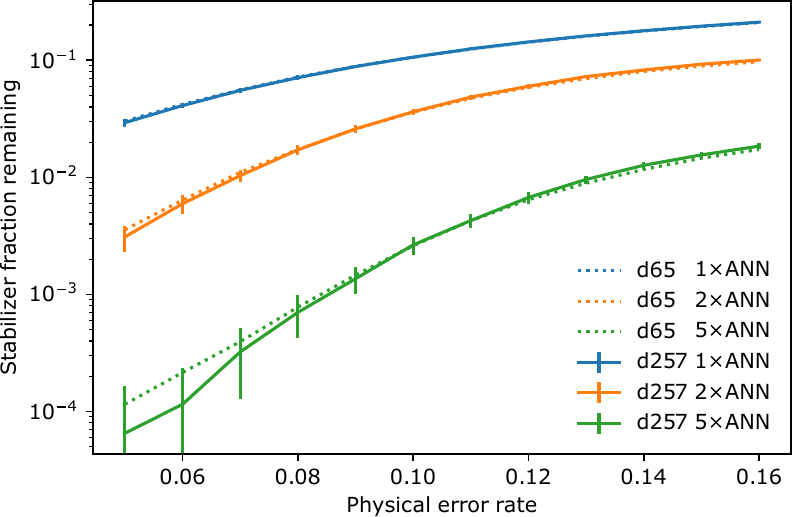}}
\end{center}
\caption{\textbf{ANN decoder syndrome processing.} Fraction of \( -1 \) stabilizer measurements remaining as a function of error rate and ANN passes -- where n$\times$ANN indicates n passes. Calculations are shown based on distances 65 and 257 square shaped surface code, with each data point calculated as the average fraction of \( -1 \) stabilizers remaining of at least 100 random depolarizing error configurations. The ANN performance is nearly independent of code distance. The error bars are shown only for the case of distance 257 and correspond to one standard deviation. The error bars for distance 65 exhibit similar trend but comparatively larger spread.}
\label{fig:4}
\end{figure}

\subsection{Decoder Training}
The ANN decoder was trained on randomly generated surface code boards. During the training process, distance 33 surface code boards of a square shape were used with square double Z cut qubits placed at random. Qubits of each board then have a depolarizing channel applied, where Pauli X, Y or Z errors occur on data qubits with equal probability \(p/3\). The stabilizers which would invert their values can then be calculated. This combination of error syndrome and boundary information serves as the input to the network during training, see for example Appendix Figure \ref{fig:input_example}. As the ANN is trained based on supervised learning, the preparation of adequate target outputs (classification labels for a given training instance) is described in detail in Appendix \ref{Appendix Neural Decoder Training}. The final training data consisted of approximately 50 million distance 33 boards, which consisted of randomly positioned distance 8 defect logical qubits. The training of the ANN decoder was performed at multiple stages. In the first stage, the training was based on small error rates (1-3\%), whereas the second stage training involved much higher error rates (9-10\%). In the final stage of training, the residual errors after testing (ANN passes) were fed back into the training data. The loss function used was binary cross entropy at initial training stages, changing to mean squared error at later stages. 
\\ \\
\noindent
As with any neural network training problem, a large predictor of the performance of the network on test data is the degree of overlap between the distribution of training data and distribution of test data. Note that this doesn't mean that performance is necessarily poor if a particular test input is absent from the training data. However, if the distribution that generated the test data is fundamentally different from the distribution that generated the training data then unexpected or undesirable behaviour may occur at inference. The behaviour of the trained ANN decoder was tested for a set of surface code boards consisting of error distributions not necessarily included in the training data sets. The proportion of the syndrome decoded by the ANN was investigated as the number of ANN passes varied. Figure \ref{fig:4} shows the performance of our ANN syndrome decoder on distances 65 and 257 square shaped surface codes. The graph plots the fraction of the original syndromes remaining after a given number of passes through the ANN (n$\times$ANN, where n is 1, 2 and 5). It can be seen that the majority of \( -1 \) stabilizer values are decoded by a single pass of the ANN at all of the investigated physical qubit error rates. At the lower physical qubit error rates this becomes more evident, with five ANN passes at an error rate of 5\% leaving only about $0.01\%$ of the original \( -1 \) stabilizers to be resolved by the mop-up decoder. This suggests promising improvements in speed, as the formation of clusters in the HDRG decoder has a timing that scales linearly with the number of \( -1 \) stabilizers. However, this does not necessarily quantify the degree to which accurate decoding is progressing because ANN corrections could be removing -1 stabilizers by completing non-trivial logical operators. This would be an example of decoding speed being gained at the expense of accuracy. When instead investigating the fraction of data qubit errors remaining as a function of physical error rate similar results are obtained when homological equivalences are accounted for (see Appendix Figure \ref{fig:target_example}). However the number of effective residual errors remaining to be decoded would also be small if corrections were applied with the simple decoder in Ref. \cite{varsamopoulos2017decoding}. The simple decoder is an important part of ``high-level" decoding, where it applies a quick set of corrections which themselves are quite likely to result in a logical error, but allow a neural network to effectively learn when this occurs for fast and accurate decoding when operating simultaneously. Indeed, full decoding results in effectively $0$, $d$ or $2d-1$ data qubit errors when decoding results in a logical $I$ operation, either of the logical $X$ or $Z$ operations, or a logical $Y$ operation respectively. No scalable method is known for training such decoders for distances beyond approximately 10 and investigations regarding the potential to use similar simple decoders was left for future work. A good measure of the value gained from applying particular corrections in surface code decoding is yet to be agreed upon. The development of such a metric would offer significant insights in the evaluation of decoding methods of systems of cooperating entities, and training of machine learning based decoders. Nevertheless combined decoding accuracy can still be effectively quantified in terms of logical error rates and thresholds, which will be discussed in the next section.
\\ \\
\noindent
Finally, we performed a rigorous end-to-end testing of our complete decoder implementation (ANN+HDRG) for many different code distances, shapes, error profiles and error probabilities. In each case, the decoder was able to accurately process all syndrome measurements, confirming a highly reliable and robust working for the chosen noise model. The Appendix Figures \ref{fig:257_12_initial}-\ref{fig:mXXX_2} show the working of our decoder through several examples which exhibit different steps during the syndrome decoding process. 
\subsection{Decoder Performance}
\begin{figure*}[htp]
\begin{center}
\resizebox{160mm}{!}{\includegraphics{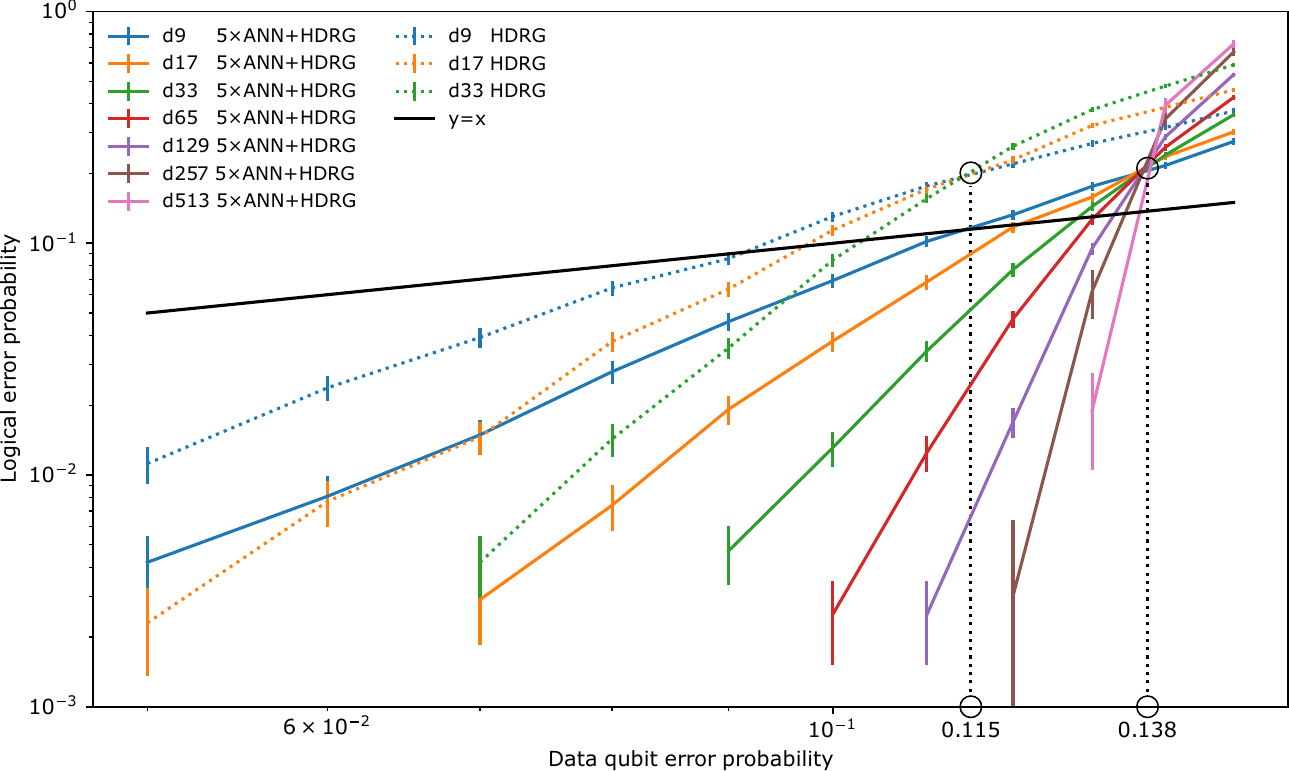}}
\end{center}
\caption{\textbf{ANN decoder logical threshold.} Logical error probabilities after decoding standard square surface codes subject to depolarizing noise on data qubits. Dotted lines show HDRG decoder performance and solid lines show performance of first passing the syndrome through 5 passes of the ANN decoder before decoding remaining syndromes with the HDRG decoder. Error bars correspond to a two standard deviation binomial normal confidence interval.
}\label{fig:5}
\end{figure*}
\noindent
Desirable characteristics of syndrome decoding algorithms include high thresholds, low logical error rates, adaptability to complex noise models commensurate with the quantum computer and low latency execution. High thresholds would allow the code to be practical on architectures with noisier qubits and gates. Low logical error rates would allow the code to be used at lower distance, reducing the number of physical qubits needed. Adaptability to the actual error characteristics of the quantum device would allow architectures with inhomogeneous and biased noise to not suffer as severe penalties in terms of logical error rate. Finally, low latency execution would allow the decoder to keep pace with the syndrome measurement process to avoid the need to slow down the measurement operations or process a backlog of syndrome data \cite{holmes2020nisq+}. It has also been indicated that such rapid application of corrections (active feedback) commensurate with measured syndrome data in practice results in lower logical error rates \cite{Andersennpj2019}. In this section, we analyse the performance of our ANN decoder and show that it offers highly promising characteristics such as reasonably high threshold coupled with low latency execution even at code distances exceeding 1000 and adaptability to hardware by processing sophisticated noise models such as spatially inhomogeneous and biased noise.
\\ \\
\noindent
We first focus on surface codes suffering from depolarising noise model and systematically analyse the performance of our ANN decoder based on two most widely adopted metrics: threshold and operation time.  Although not the focus of this study, performance at smaller distances is relevant for quantum devices currently being developed. Low error rate effects, such as decoding codes with small numbers of errors are also more readily investigated for smaller codes. In this regime, local decoders are susceptible to make mistakes on syndromes which would be otherwise correctable with a global perspective. This can cause an increase in logical error rates, unless potential safety measures are put in place to avoid any avoidable increases in logical error rates. Further calculations suggest that the relevance of some of these effects quickly diminished as larger error rates and code distances were investigated (see Appendix Figure \ref{fig:binom_error_rates}). We have instead chosen to focus our investigation on decoding for systems at larger scales, approaching those commensurate with the needs of multiple logical qubits performing logical operations during a fault tolerant quantum computation \cite{gidney2019factor}. Figure~\ref{fig:5} shows the decoding accuracy results. The calculations were performed for square shaped surface codes of distances between 9 and 513. Dashed lines correspond to performance when using the non-ML HDRG decoder alone. Solid lines correspond to the decoding performance after five rounds of ANN processing (5$\times$ANN) followed by mop-up step based on HDRG decoder to process any residual syndromes (see Appendix Figure \ref{fig:just_flowchart} for details). We note that the use of the ANN decoder improves the threshold to approximately 13.8\%, when compared to 11.5\% threshold from non-ML HDRG decoder. We attribute the higher threshold of the ANN+HDRG decoder to its ability to deduce sites of Y errors with high probability, as it decodes X and Z syndromes together compared to the HDRG decoder alone which decodes X and Z syndromes separately. Overall, the threshold of our ML-based decoder is comparable to other ML-based decoders, which have reported thresholds in the range of 10 to 16\% for noise models ranging from bit-flip noise to depolarizing noise \cite{Meinerz2021arxiv}. The non-ML based decoders have achieved somewhat higher thresholds, reaching 18\%, however, these high threshold were reported at the expense of scalability (code distances limited to 100). In our ANN decoder construction, the primary focus has been on the scalability to large distances and operational speed, demonstrating the performance for code distances above 1000 while achieving an adequate threshold of 13.8\%. We also note that the slightly lower threshold of our ANN based decoder can be attributed to its overzealous correction strategy based on the information found in the window of local observation. In principle, the window of local observation can be increased, which after rigorous training should allow substantial increase above 13.8\% threshold.
\begin{figure*}[htp]
\begin{center}
\resizebox{170mm}{!}{\includegraphics{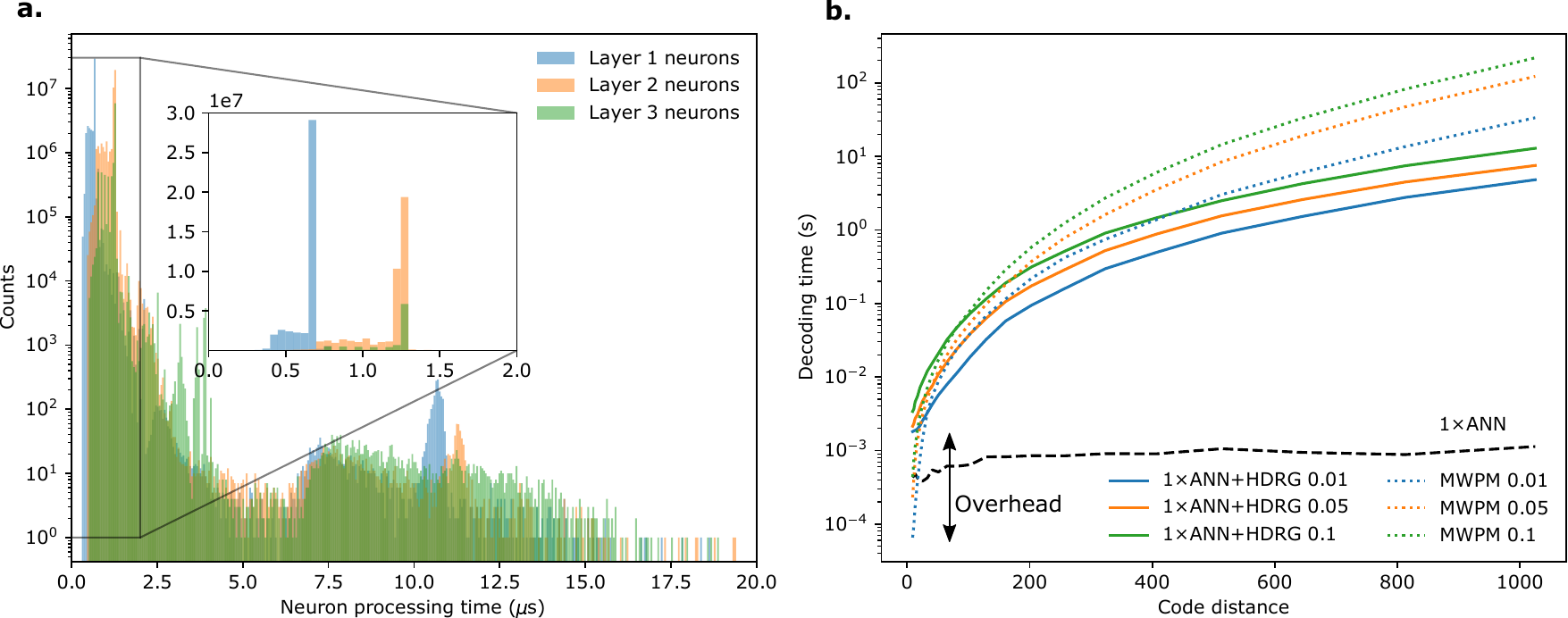}}
\end{center}
\caption{ \textbf{ANN decoder operation time.} \textbf{a.} Histograms are plotted to show the distribution of counts over time taken to compute individual neurons for the trained neural decoder when implemented in C++. The inset shows the distribution with a linear scale for times less than two microseconds. Bin width is set to 50 nsec. \textbf{b.} Decoding time of the ANN+HDRG decoder (only 1 ANN pass followed by HDRG) as a function of code distance. The overall time is dominated by HDRG mop-up step and therefore multiple ANN passes will have negligible impact on overall time. Different colours indicate depolarizing error rates of 1\%, 5\%, and 10\%. The corresponding operation times of MWPM with the PyMatching package \cite{Higgott2021arxiv} are also shown for comparison. The implemented ANN+HDRG decoder scales better than the MWPM implementation with code distance, and can be further improved by dedicated hardware allocation (overhead reduction).  For reference, the total time taken in one pass through the ANN is also shown (broken black line), which is independent of error rate and code distance.
}\label{fig:6}
\end{figure*}
\noindent
\\ \\
\noindent
The second important performance parameter is the speed of decoding operation. To investigate performance of individual components of the computation, the neural network was written in C++ code, where only the evaluation of the neural network function was implemented. This reduces the computational overhead which arises in implementations which also include methods of general network construction and training. Timings to compute convolutional neurons were measured for each layer of the network on a distance 33 code and the results are shown in Figure~\ref{fig:6} (a). This was calculated for a non-parallel implementation of the neural decoder. Multiple peaks can be associated possibly with cache misses during the computation. Each neuron in layer 1 gets computed on average slightly faster than neurons in layers 2 and 3, which take approximately the same amount of time on average. This is consistent with the convolutional kernel being over 4 channels for the first hidden layer and over 9 channels for the second and third hidden layers. The majority of neurons had a computation time beneath approximately 1.5 $\rm \mu$sec, which can be taken as an upper bound for the time taken to compute a full convolutional layer of the ANN if implemented with parallel computation of neurons, possibly with dedicated hardware such as FPGAs or ASICs. The applicability of this approach in practice relies on the availability of such classical computing hardware adjacent to the quantum device. Fully parallelizing over each data qubit would hence require the ability to perform O($d^2$) simultaneous kernel evaluations. Under these assumptions, the time to compute the two hidden layers and final output layer would be 4.5 $\rm \mu$sec, which is beneath the required reaction time estimated as 10 $\rm \mu$sec to perform Shor's algorithm capable of breaking the 2048-bit RSA cryptography \cite{gidney2019factor}.
\\ \\
\noindent
Full decoding time (ANN+HDRG) as a function of code distance is shown in Figure~\ref{fig:6} (b). The operation times of decoder implementations were measured on a system with 12 CPU cores and 1 GPU (see Appendix \ref{Appendix Hardware and Software Details} for more details) with the ANN decoder written using Tensorflow~\cite{tensorflow2015-whitepaper} functions to reduce inference latency. One pass was taken through the neural network and a HDRG decoder was used to correct any residual errors. The results displayed in Figure \ref{fig:6} (b) directly compare with PyMatching's \cite{Higgott2021arxiv} implementation of MWPM tested with the same set-up (with locality parameter set to the default value of 30). MWPM decoding has been the standard benchmark against which other decoder candidates are compared. PyMatching is an implementation with an average time complexity of O($d^{2.1}$)\cite{Higgott2021arxiv}. An alternative choice for comparison is the Union-Find decoder with complexity O($d^2\alpha({d^2})$) where $\alpha$ is the inverse of Ackermann’s function \cite{delfosse2021almost}, which is close to scaling quadratically with distance. PyMatching was chosen as it gives an indication of performance of the well studied MWPM decoder when implemented in Python.  We note that decoding times increase as a function of surface code distance as well as with respect of error rate in all cases. This is a result of the size of the graph problem increasing as the number of \( -1 \) stabilizer values increase. However, this is not the case for the ANN decoding step. The ANN computes the same function, regardless of how many \( -1 \) stabilizer measurements are present, and hence has an error rate independent execution time (broken line plot). However, as the error rate is increased more passes may be necessary to process a sufficient number of \( -1 \) stabilizers (Figure~\ref{fig:4}). For larger distances the operation time of multiple ANN passes will have negligible contribution to the overall decoding time.  The PyMatching implementation has comparable decoding times at code distances below approximately 100 -- for code distances above 100, our ANN+HDRG decoder implementation has better decoding times. However, the total decoding time in this regime is still tens to hundreds of milliseconds, which is still too long to meet the demands of coherence times of some proposed surface code architectures, such as superconducting qubits. Other surface code architectures do meet demands for longer coherence times \cite{kobayashi2021engineering}, however it is possible to significantly reduce total execution time with further optimizations, such as the use of dedicated hardware. At code distances above 1000, which would be commensurate with scales of fault-tolerant quantum information processing featuring numbers of logical qubits relevant for practical applications \cite{gidney2019factor}, we estimate that our ML-based decoder implementation offers roughly two orders of magnitude better operational times than the PyMatching algorithm. The better performance of our implementation can be attributed to the ability of ANNs to efficiently make use of parallel processing to offer a significantly smaller decoding problem to the ``mop-up'' decoder, as described during decoder construction. This can be expected when other decoders based on matching are used in the mop-up step, such as MWPM (Appendix Figure \ref{fig:mwpm_mopup}). Importantly, we point out that the operation time of our ANN decoder can be drastically improved by reducing overhead which indicates very slow performance even at trivially small code distances. We expect that by exploiting standard parallel CPU and GPU implementations of the neural network could offer $\leq$ 10 $\rm \mu$sec target decoding time. Also, improvements to ``mop-up'' decoder latency (overhead) would significantly reduce total decoding time as it is the main bottleneck in the current implementation. 
\\ \\
\noindent
After benchmarking the performance (scalability and low latency) of our ANN+HDRG decoder for the depolarising noise model, we now turn our attention to the third key desirable property, which is adaptability to hardware. Quantum computers are still under development and the exact nature of noise in a scalable quantum device is still an open question. It has been reported that qubit dephasing noise (Z errors) is the dominant source of noise in quantum hardware \cite{Ataides_NComms_2021}, therefore a desirable property of surface code decoders is to handle noise which is biased towards Z errors. Indeed our decoder can handle a variety of complex noise profiles, which include spatially inhomogeneous errors and biased noise without the need for any re-training or special treatment. Appendix Figures \ref{fig:inhom_2_2_flowchart1}-\ref{fig:biased_20_20_60_flowchart4} show a few examples of surface codes with distance 257 with various configurations of biased noise and inhomogeneous noise, which are processed by our decoder without the loss of performance.
\\ \\
\noindent
A fault-tolerant quantum algorithm could be implemented through defects, braids and/or lattice surgery. Recent work has reported that lattice surgery could significantly reduce hardware resources \cite{Fowler_lowoverhead_2019}, and therefore another desirable feature for surface code decoders working with future fault-tolerant quantum hardware is to process surface codes with arbitrary size and shapes which are fundamental to implement lattice surgery structures. Our ANN decoder is capable of processing surface codes structures relevant for defects, lattice surgery and/or braiding operations, marking an important step towards implementation of a fault tolerant surface code decoder applicable to experimental devices. This is due to the innovative approach underpinning our ANN decoder construction which involves explicit knowledge of boundaries stored in additional channels of the input layer and progressively understood in the training process. This is also a significant advancement over previously reported ML-based decoders in the literature which do not offer such capabilities. Explicit examples of rotated surface codes and surface codes with non-trivial boundaries are included in the Appendix Figures \ref{fig:rot_12_flowchart1}-\ref{fig:mXXX_2}.
\section{Summary and Outlook}
This work reports a scalable and fast decoder implementation for surface code syndrome processing, which exploits the computational power of a machine learning technique. The decoder performance has been analysed for surface codes with distances up to 1025 (over 4 million physical qubits), which is the largest such demonstration for an ML-based decoder to-date. The decoder has demonstrated a logical error threshold of about 13.6\% which is comparable to non-ML and ML based decoder previously reported in the literature. The decoding accuracy (threshold) could be further increased by using additional hidden layers or otherwise increasing the window of local observation for each data qubit. The main source of the slightly smaller threshold of our ANN decoder compared to other schemes such as fully connected (dense) ANNs and MWPM is attributed to its overzealous corrections made considering the information found in the window of local observation. Thus, if the window of local observation is increased, it is expected that after rigorous additional training, the decoding accuracy will be comparable to any state-of-the-art non-ML algorithm such as MWPM. 
\\ \\
\noindent
The key novelties of the ANN decoder developed in our work include its scalability to systems of many logical defect qubits, compatibility with structures of lattice surgery, and fast operation times in particular based on a future dedicated hardware implementation. The operation time of our ANN decoder is O(1) with respect to code distance, and the overall ANN+HDRG operation time outperforms optimised implementation of the MWPM algorithm at large code distances ($\geq$ 1000) relevant for fault-tolerant quantum information processing. However, the total decoding time is nevertheless limited by the time taken for mop-up decoding. This leads to total decoding time complexity which is not O(1). Further investigations may be possible to reduce the overhead caused by the mop-up component of a hybrid decoding strategy. Previous work on neural networks, outside the context of surface codes, when optimized and run on dedicated hardware has already demonstrated that inference latencies beneath 1$\mu$sec are achievable and further optimisation is expected to reach inference latencies beneath 0.1$\mu$sec \cite{nikonov2019benchmarking}. However, such performance was demonstrated on 32x32 single channel inputs with convolutional kernels of width 5 and remains to be demonstrated to be able to persist for larger input sizes and kernel widths. Under these assumptions, this suggests that, in principle, our ANN decoder could perform decoding operation in 0.1$\mu$sec time-scale which is commensurate with the current hardware coherence limitations. Additionally, the need to use a non-ML mop-up decoder such as HDRG algorithm employed in our work makes the latency of decoding depend on the execution time of the global decoder, limiting the overall speed of operation. Future subsequent work will explore a faster implementation of a mop-up algorithm based on renormalization techniques and/or further optimisation of local decoding within the ANN decoder construction that can be implemented to eradicate the need for a mop-up step. Another interesting approach under consideration would be to construct a scalable fully convolutional compatible method of renormalization similar to the approach discussed in Ref. \cite{ni2020neural}.
\\ \\
\noindent
We note that our work demonstrates processing of a variety of noise modes including depolarising error model, spatially inhomogeneous noise and biased noise models. This is a significant advancement over the bulk of the existing ML-based surface code decoding literature, which has primarily investigated either bit-flip errors or depolarising errors (see Table \ref{review_table} in the Appendix). However, in an experimental set-up, it is anticipated that ancilla qubit (syndrome) measurements may also experience errors which are only scarcely considered in ML based decoders to-date. Our ANN decoder could possibly be applied on each round of syndrome changes individually, perhaps with some additional training. It would leave most individual changed stabilizers from ancilla measurement errors to be decoded by the mop-up decoder, due to a lack of justification supporting the application of large correction chains. However, a generalization directly compatible with chains of errors across multiple cycles would likely perform better. Preliminary work on neural network decoders including measurement errors, as well as general gate errors, has been reported that take multiple rounds of syndrome measurements as input \cite{baireuther2018machine}. Our decoder based on fully convolutional construction is compatible to deal with such syndrome measurement error models, and future work will extend the established decoder by constructing 3D convolutional layers or convolutional LSTM layers to accommodate such error models. Another aspect is the use of an explicit error model, which itself causes the knowledge of an adequate error model a prerequisite to the application of any decoding strategy in practice. As quantum devices are currently at small-scale, the actual nature of noise in a scalable quantum hardware which is still 5-10 years into future is not yet clear. It is possible that the errors present in real devices will experience spatial and/or temporal correlations, and may depend on current and previous processes that have occurred within the vicinity of a qubit. A salient feature of our ANN decoder is its construction based on multi-channel input layer which provides flexibility to add additional channels to tackle complex noise models which might become available as the quantum hardware becomes more mature. Thus, methods of determining beneficial local changes to the decoding strategy will be investigated. This will be aided by training on a real quantum device.
\\ \\
\noindent
Overall, our work has achieved an important milestone towards the realisation of fault-tolerant quantum computing by designing a scalable and fast ML-based surface code syndrome decoder whose experimental implementation could enable elusive quantum advantage for real-world applications.
% \noindent
% % \\ \\
% \normalsize
\noindent
\\ \\
\noindent
\textbf{Data Availability:}
The data that support the findings of this study are available within the article and the Appendix. Further information can be provided upon reasonable request to the corresponding author. 
\normalsize
\noindent
\\ \\
\noindent
\textbf{Code Availability:}
The source code used to generate figures in this work can be provided upon reasonable request to the corresponding author. 
\noindent
\\ \\
\noindent
\textbf{Acknowledgements:} This work was supported by the Australian Research Council funded Center for Quantum Computation and Communication Technology (CE170100012). Assistance of resources and services were provided from the National Computational Infrastructure, which is supported by the Australian Government. This research was undertaken using the LIEF HPC-GPGPU Facility hosted at the University of Melbourne. This Facility was established with the assistance of LIEF Grant LE170100200.
\\ \\
\noindent
\textbf{Author Contributions:} M.U. and L.C.L.H. planned and supervised the project. S.G. developed the artificial neural network framework with input from M.U. and L.C.L.H. S.G. carried out the simulations. All authors contributed to the analysis of data. M.U. and S.G wrote the manuscript with input from L.C.L.H.
\\ \\
\noindent
\textbf{Competing Financial Interests:} The authors declare no competing financial or non-financial interests.

\bibliographystyle{quantum}
% \bibliography{sample}

\newpage
\onecolumn

\appendix
\section{Noise Models}
\noindent
\noindent
The noise present in real quantum devices can be approximated with the use of noise models. While general noise models can be difficult to simulate efficiently, particular instances such as stochastic Pauli errors can allow efficient classical simulation. The noise models discussed in this work are described below.
\subsection{Depolarizing Noise}
\noindent
When qubits experience depolarizing noise they have an $X$, $Y$ or $Z$ error applied with equal probability. In this work, depolarizing noise models are defined by a single parameter, $p$. With probability $p$, each qubit is is independently chosen to experience an error. The type of error is chosen uniformly from the set $\{X, Y, Z\}$. Note that the identity error, $I$, which is often present in other definitions of depolarizing noise, is absent from this set.
\subsection{Biased Noise}
\noindent
Biased noise was defined similarly to depolarizing noise, except with the condition of equal probability $X$, $Y$ or $Z$ errors lifted. Similar to depolarizing noise, $p$ describes the probability that each qubit independently experiences an error. Unlike depolarizing noise, however, the type of error to occur is $X$ with probability $p_X$, $Y$ with probability $p_Y$ and $Z$ with probability $p_Z$. Note that again the identity error is not included as a possible error. Hence, the equation $p_X+p_Y+p_Z=1$ must be satisfied. 

\subsection{Spatially Inhomogeneous Noise}
\noindent
Spatially inhomogeneous noise models were defined with independent error probabilities, $p_0, p_1, ..., p_{N-1}$, assigned to each of the $N$ data qubits to specify the probability that each experience an error. The spatially inhomogeneous noise model can be also chosen to be biased. In this case, each data qubit also has specified a probability of $X$ error as $p_{i,X}$, probability of $Y$ error as $p_{i,Y}$ and probability of $Z$ error as $p_{i,Z}$, where $i$ takes values $0, 1, ..., N-1$ for N data qubits. As the identity error is not considered, the equation $p_{i,X}+p_{i,Y}+p_{i,Z}=1$ must be satisfied for each qubit.  
\noindent

\section{A  Brief Literature Survey of non-ML and ML based Syndrome Decoders}\label{Appendix Literature Survey}
\noindent
During surface code operation, a crucial step is the processing of measured stabiliser syndromes leading to efficient formulation of corrections to be applied to data qubits. This process is performed by classical algorithms known as decoders. Over the years, significant search efforts have been invested in the design of efficient decoders. Historically, the decoder algorithms have been based on graph matching problems, without involving any machine learning technique, which we shall call here as non-ML decoders. Example's include minimum weight perfect matching (MWPM) decoding \cite{fowler2010surface, fowler2012towards, fowler2013optimal}, related matching/clustering based decoders such as renormalization group decoding \cite{watson2015fast, duclos2010fast} and cellular automaton decoding \cite{holmes2020nisq+}. MWPM decoding operates by choosing a correction operator consistent with the most probable error (under an independent Pauli X and Z error error model). This is can be done in polynomial time by calculating the solution to a graph problem using the Blossom algorithm. Further accuracy can be achieved by performing additional matchings, on graphs re-weighted based on the results of previous matchings in order to better correct correlated errors \cite{fowler2013optimal}. MWPM has had substantial development and its adequate accuracy allows it to be considered the benchmark which all other decoding algorithms are compared to. It is unknown whether it can be executed with latencies that can keep pace with expected syndrome measurement times of the fastest quantum computing architectures. Hard decision renormalization group decoding \cite{watson2015fast} is expected to achieve better latencies by utilizing iterative methods at different scales. It generally achieves lower thresholds and accuracies than MWPM. Cellular automaton decoders \cite{holmes2020nisq+} consist of repeating blocks of identical cells which through communication with one another decide on correction operations based on predefined policies. They generally achieve the lowest latency execution for small error rates but begin to slow down at larger error rates, with communication time between cells being the main issue. There have been suggestions to construct cellular automata to execute MWPM algorithms \cite{fowler2013minimum}, however this technique remains an area of work to achieve a robust proof of concept implementation. 
\noindent
\\ \\
\noindent
The application of machine learning techniques for the design of surface code decoders is a relatively new area of research, with most of the studies appearing within the last five years. Recent research has reported the design of artificial neural network (ANN) syndrome decoders \cite{torlai2017neural, krastanov2017deep, varsamopoulos2017decoding, baireuther2018machine}, with preliminary studies demonstrating that ANN decoders could offer highly promising performance for small surface codes. Many varieties of neural networks have attempted to be trained to act as good surface code decoders, which include dense networks \cite{varsamopoulos2017decoding}, networks with convolutional layers followed by dense layers \cite{davaasuren2020general}, and network utilizing long short-term memory (LSTM) layers \cite{baireuther2018machine}. A comprehensive survey of ML-based decoders along with performance metrics such as threshold, maximum distance demonstration, ML technique used and noise models is provided in Table S1.

\section{Introduction to Artificial Neural Networks}\label{Appendix Introduction to Artificial Neural Networks}
For general audiences, a short overview of artificial neural networks will be discussed here, providing sufficient background for the understanding of the neural network decoder structure developed and discussed in this work. A more in depth overview of current approaches utilizing artificial neural networks in scientifically related problems is presented in Ref. \cite{dunjko2018machine}. 
\noindent
\\ \\
\noindent
Artificial neural networks are composite functions which have structure built upon individual objects called neurons. A neuron can be defined in terms of the connections which make up its input together with their associated weights, as well as a single bias and activation function. The behaviour of a single neuron can be described as,

\begin{equation}\label{eq:node_execution}
    Z=A\left( \sum_{k} w_k Y_{k}+b \right),
\end{equation}
\noindent
\\ \\
\noindent
where \(Z\) is the value of the neuron of interest, \( w_k\) is the weight between the neuron of interest and the $k^{th}$ neuron of the previous layer, \( Y_k\) is the value of the $k^{th}$ neuron of the previous layer (or the $k^{th}$ input of the neural network), \(b\) is the bias of the neuron of interest and \(A\) is the activation of the neuron of interest.
\noindent
\\ \\
\noindent
Neural networks are often shown with sets of neurons drawn together when they possess together an associated set of inputs. These are known as layers in a neural network. Examples include dense layers which are comprised of neuron with connections to every neuron of a previous layer and locally connected layers which are comprised of neurons with connections to small neighbourhoods of the previous layer. The value of a set of neurons at a particular layer can be written as a function of the neural network inputs by recursively applying equation \ref{eq:node_execution}. The final layer of neurons are considered the neural network output.
\noindent
\\ \\
\noindent
The weights and biases of the neural network are adjusted in order to achieve adequate performance for a particular learning task. This is usually done based on gradient update methods computed from a loss function. Methods based on gradient descent are possible due to the care taken to ensure that the neural network output is a differential function of the weights. In a supervised learning setting, where the target behaviour is defined explicitly for a set of inputs, an example of a cost function is the sum of square errors,

\begin{equation}
    L(y, \hat{y})=\sum_{i} (y_i- \hat{y_i})^2, 
\end{equation}
where \( y \) is the neural network output and \( \hat{y}\) is the desired output for a batch of inputs, and where their subscripts represent particular inputs of that batch.
\noindent
\\ \\
\noindent
Other parameters which are not deduced through training on data are known as hyperparameters. These include the neural network connectivity structure, activation functions and training details such as batch size, loss function and gradient update rule.
\noindent
\section{Neural Decoder Construction}\label{Appendix Neural Decoder Construction}
A syndrome decoder processes error information and suggests appropriate corrections. This is equivalent to calculating the most probable equivalence class of error to have occurred, which is consistent with each measured syndrome. This can be written mathematically as,
\begin{equation}
    D_{\mathrm{opt}} (\overrightarrow{s}) =\underset{E}{\arg\max} (P(E)),
\end{equation}
\noindent
where \( E\) represent the equivalence classes of possible correction operations. Calculating the most probable equivalence class directly is difficult in general as performing the calculation directly requires taking into account the occurrence probability of each error consistent with the observed syndrome. The number of errors grows exponentially with the number of qubits in the system for a given error probability. Even approximating the occurrence probability of each equivalence class is difficult when performed directly as it would require unbiased sampling from the set of errors consistent with the syndrome. Any correction operation of the most probable equivalence class would work just as well if no gate errors are present (as was assumed in the rest of this paper). If gate errors were present then that should be kept in mind when choosing the most appropriate method of correction of the most probable equivalence class. Small optimizations would be possible by performing correction operations using qubits with the most accurate gates.
\\ \\
\noindent
The neural network decoder implemented in our work consists entirely of locally connected neural network layers. This is in contrast to previous decoders reported in the literature, which also make use of convolutional layers, but have dense final layers \cite{davaasuren2020general}. Our decoder is inspired from pattern matching and cellular automaton based approaches which correct data qubit errors based on local observations of the syndrome. Also of inspiration was the notion of making use of the translational symmetry present in surface codes, an idea continually discussed since the first mention of machine learning decoders \cite{torlai2017neural}. Local observation  alone, without communication, is unable to result in a decoder with a nonzero threshold for the 2D surface code. This is because for any code decoder with fixed radius of local observation, there exists a distance for which otherwise correctable errors become uncorrectable. This occurs when the errors form chains of length larger than the radius of local observation. The approach presented here approaches this problem by propagating information between convolutional layers, increasing the local receptive field and allowing correction of clusters at small scales. At large scales a matching based decoder is used to ``mop-up'' any residual errors that remain of sizes beyond the local receptive field. The residual errors that remain are at a much smaller quantity and this is used to achieve an overall speed up in decoding compared to using matching algorithms alone. More than a single pass through the ANN decoder may be optimal to minimize the total computation time. This should be calibrated by investigating the ``mop-up'' speed-up achieved as a function of syndrome sparsity. A flowchart of the decoding process is presented in Appendix Figure \ref{fig:just_flowchart}.
\noindent
\\ \\
\noindent
The fully convolutional decoder implemented here uses a four channel input layer where the first two channels correspond to the X and Z syndrome information and the last two channels correspond to the X and Z boundary qubit information. Boundary qubits are defined as data qubits which when acted upon by an X or Z errors cause exactly 1 stabilizer generator to change values. This is done to give sufficient information to identify appropriate correction operations given the local syndrome environment and local boundaries of the code. The inclusion of boundary information allows the decoder to be trained on and applied on surface codes of varying shapes and sizes.
\noindent
\\ \\
\noindent
The next two layers of the surface code, the hidden layers, are convolutional layers. The use of convolutional layers allows the first hidden layer to be interpreted to consist of a multi-channel image containing local information about the syndrome and boundary environment. The locality of the information is defined by the size of the convolutional kernels. Larger convolutional kernels result in information being able to be transmitted about the presence of larger clusters. An alternate way to achieve the ability to identify larger clusters is to use additional hidden layers. An effect resembling a light-cone occurs where the deeper hidden layers consist of neurons which depend on a larger window of local observation, shown in Appendix Figure \ref{fig:decoder_layers} (b).
\noindent
\\ \\
\noindent
For simplicity, we chose all the hidden layers to contain the same shape and number of kernels. The kernel size of 11x11 (shaped like a distance 6 surface code) was chosen to be able to identify small clusters of errors (2-3 errors) as well as information about adjacent clusters and boundaries in the first hidden layer. Each convolutional layer consists of a given number of convolutional kernels, with each kernel forming a channel of the hidden layer. These kernels/channels can be thought of as feature maps and the number of kernels needed can be estimated by counting common features that appear in the training data. We counted the 6 unique syndrome patterns corresponding to X, Y and Z errors in the bulk (see Figure \ref{fig:2}) and assigned 3 more to identify errors near boundaries. The identification of features corresponding to more complex syndrome patterns was left to subsequent layers. Of course, the actual form the features take is determined during training and may show no resemblance to the features intuitive to human beings. These arguments can only give reasonable initial ranges for parameters and more rigorous arguments/direct search is needed to find optimal hyperparameter values.
\noindent
\\ \\
\noindent
The output layer is again another convolutional layer with two kernels being trained. Each kernel results in a channel of the layer and thus the output layer will consist of two channels. These two channels are interpreted to contain the probabilities of the associated qubits requiring an X or Z correction operation. As the output will be interpreted as a probability, an activation function such as a sigmoid is used to ensure that the output is between zero and one. Finally, all the data qubits with a probability larger than 0.5 of being in error are recommended to have the appropriate correction operations applied. This is in contrast to some previous neural decoders suggested in the literature, which instead encourage sampling from the distribution of the output until a set of errors is found which is exactly consistent with the syndrome measured \cite{krastanov2017deep}. Such a selection processes requires a number of samples which grows exponentially with the number of data qubits of the system and hence is not considered in this work which aimed at decoding boards at distance beyond 10.
\noindent
\\ \\
\noindent
Thus the network can be interpreted as an image to image network where the input image consists of four channels and the output image consists of two channels (see Appendix Figure \ref{fig:decoder_layers}). The output layer consisting of a sigmoid activation function leads to the behaviour of this network easily understood as qubit-wise error state logistic regression. All prior layers use ReLU activation functions. The use of exclusively convolutional layers allows the network to operate on surface code boards of arbitrary shape and size, so long as the input channels are constructed to adequately convey the required structural information. Two hidden layers were used, each containing 9 channels, and all kernels were 11x11 in size. The use of 3 convolutional layers results in a local receptive field with a radius of 15 cells.
\noindent
\section{Neural Decoder Training}\label{Appendix Neural Decoder Training}

\begin{figure*}[htp]
\begin{center}
\resizebox{170mm}{!}{\includegraphics{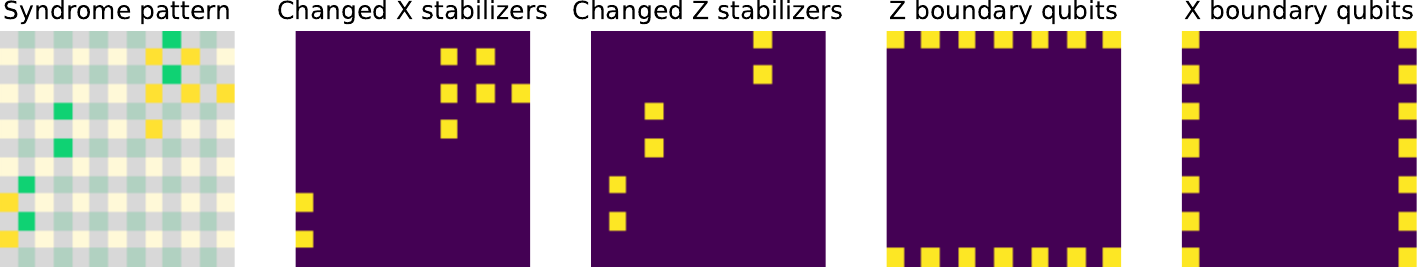}}
\end{center}
\caption{
    The channels for neural network input corresponding to a syndrome pattern of a distance 7 surface code.
}\label{fig:input_example}
\end{figure*}

The decoder is trained on randomly generated surface code boards. Distance 33 surface code boards of a square shape were used with square double Z cut qubits placed at random. Qubits of each board then have a depolarizing channel applied, where Pauli X, Y or Z errors occur on data qubits with equal probability \(p/3\). The stabilizers which would invert their values can then be calculated. This combination of error syndrome and boundary information is then the input to the network during training, see for example Appendix Figure \ref{fig:input_example}. As the training is based on the supervised learning principle, appropriate classification labels must also be generated which associate a given instance of extracted syndrome with appropriate corrections to be applied. We refer these classification labels as target outputs. 
\noindent
\\ \\
\noindent
Defining adequate target outputs is the most non-trivial part of training. The past literature has used the set of random errors, which generated the set of inverted stabilizers as the target output \cite{krastanov2017deep}. Such networks will never converge to optimal deterministic decoding performance. This is because the homologically equivalent errors present during training which occur at equal likelihood cause network outputs to favour taking outputs of 0.5 on the qubits which have unequal variable error states. Thus for most non-trivial errors the network will consistently output values close to 0.5 near \( -1 \) stabilizers. To circumvent this issue we choose to instead make the target output instead homologically equivalent to the errors which generated the changed stabilizers.
\noindent
\\ \\
\noindent
Using a target output which is homologically equivalent to the errors which generated the changed stabilizers solves many problems. Firstly, it permits the use of a unique error to be used as a target for each homologically distinct equivalence class for each possible syndrome measurement. As a consequence it also solves the problem which causes network uncertainty for syndrome measurements which have many equally likely generating errors. Lastly it allows the output of the network to be interpreted as a specification of an equivalence class rather than merely the probability a particular qubit operation may be consistent with the error syndrome observed.
\noindent
\\ \\
\noindent
However, the construction of a homologically equivalent error representative for every possible error chain is no easy task. Such a function does not have a unique definition and the adoption of a poor definition may result in incompatibilities with a local approach to decoding. For instance if such a function is demanded to always give homologically equivalent error chains of minimum weight, then such a function necessarily lacks two-fold rotational symmetry and reflection symmetry. It would also have difficulties associated with translational invariance. This is a result of the existence of error chains which have a syndrome symmetric under such transformations. It is still an open question whether a favourable function exists which facilitates efficient decoder training at all distances.
\noindent
\\ \\
\noindent
Instead, heuristics were developed to deduce favourable homologically equivalent representative errors. Firstly, if present, any full weight 4 stabilizers are removed from the generating error chain. These do nothing but increase the tendency of the network to favour outputs close to 0.5 as they would be a target correction operation that occurs with the lack of any evidence. Next, ideally all stabilizers operators within the generating errors would be removed. However, the total number of stabilizer loops to check is exponential in the number of stabilizer qubits and occur incredibly infrequently for randomly generated data. Hence larger stabilizer loops were not checked for or removed. Next we toggle stabilizer operators for which more than half of its qubits are in errors. This, again, strictly reduces the number of errors in the target set. Next, all diagonal errors are replaced to instead by in the shape of L or J. This removes the ambiguity of whether a diagonal error should be corrected by a chain of either shape L or upside down L, or shape J or shape r. Such corrections are of equal likelihood to have occurred naturally or from prior alterations to the target corrections. Lastly, errors which have syndromes at the four corners of a square and appear as parallel vertical or horizontal chains and altered to strictly be one or the other (never both for the training of the same decoder). These syndrome patterns consist of half filled stabilizers and hence occur with equal likelihood and have little overlap in the qubits they effect. Thus, unless remedied, they instil uncertainty in the decisions made by the network. Ideally all occurrences of half (or more than half) filled stabilizer operators would need to be manipulated to cause errors to only occur in a single form. However, again the number of these operators increases exponentially with increases to system size so only operators in the form of small squares were checked and manipulated (see Appendix Figure \ref{fig:target_example}).
\noindent
\\ \\
\noindent
The final training data consisted of approximately 50 million distance 33 boards, which consisted of randomly positioned distance 8 defect logical qubits. Error rates were increased from 0.01 at early training to 0.1 at late training. The loss function used was binary cross entropy and then changed to mean squared error at late training. As with any neural network training problem, a large predictor of the performance of the network on test data is the degree of overlap between the distribution of training data and distribution of test data. Note that this doesn't mean that performance is necessarily poor if a particular test input is absent from the training data. However, if the distribution that generated the test data is fundamentally different from the distribution that generated the training data then unexpected or undesirable behaviour may occur at inference.

\section{HDRG Decoder}\label{Appendix HDRG Decoder}
A HDRG decoder was written in Python to perform the final step of the decoding process. It is based off the general version described in reference \cite{watson2015fast} and adapted for the case of depolarizing noise on qubits with no measurement errors. A complete flow chart diagram of our decoder's working is provided in the Appendix Figure \ref{fig:just_flowchart}. A syndrome measurement array from a surface code set up with a random distribution of depolarising errors is first passed through the ANN decoder multiple times (see Figure~\ref{fig:3} for remaining syndrome fraction as a function of error rate) until the leftover syndrome are sparse -- typically less than 10\%. The final step in our decoding process is to pass remaining syndromes through to a HDRG mop-up decoder, which clears any residual errors, not already recognised by ANN. The HDRG decoder implementation in our work is inspired by Watson \textit{et al.} \cite{watson2015fast}. The decoder was implemented in a python program and used nominal compute resources without any particular emphasis on computational efficiency. The HDRG algorithm is based on clustering, neutral annihilation, and increasing the radius of clustering in an iterative process, until all of the changed stabilisers are resolved.
\noindent
\\ \\
\noindent
The HDRG decoder operates with the repetition of 3 steps: clustering, neutral annihilation, and renormalization. Also, an important definition is that of a delta-connected cluster of –1 stabilizers, which is defined as a set of –1 stabilizers for which each –1 stabilizer of the set is no further than delta data qubit errors away from at least one other –1 stabilizer of the set. The clustering step consists of the formation of clusters of –1 stabilizers which form delta-connected sets. The neutral annihilation step consists of applying corrections to the clusters which either have an even number of –1 stabilizers or are delta-connected to a boundary. In the case of square surface codes, it is sufficient to apply corrections to move all the –1 stabilizers of the cluster to the first stabilizer of the cluster. If the cluster was instead delta-connected to a boundary, then it is sufficient to move all the –1 stabilizers of the cluster to that same boundary. The renormalization step doubles the value of delta, which was initially set as 1. The algorithm ends when no –1 stabilizers remain. 
\noindent
\\ \\
\noindent
The version of HDRG implemented in Python forms clusters by instead looking at square neighborhoods around each –1 stabilizer for simplicity. Adjacent stabilizers then appear near the edges of the neighborhood when they are delta data qubit errors away from the –1 stabilizer of interest and near the corners of the neighborhood when they are 2xdelta data qubit errors away from the –1 stabilizer of interest. This is used to define a graph, where nodes are –1 stabilizers and edges exist between nodes which appear in their neighborhoods. The set of connected clusters are then found by running a SciPy connected\textunderscore components function to find the connected components of the graph problem. Once this is done, the neutral annihilation and renormalization method steps can proceed. 
\noindent

\noindent
\section{Further Performance Analysis}\label{Appendix Further Performance Analysis}
In the main text, the proportion of decoding performed by the ANN was shown by finding the fraction of -1 stabilizers remaining after varying numbers of ANN passes at a variety of error rates. The proportion of the decoding problem performed by the ANN can also be quantified by investigating the fraction of errors remaining after a given number of ANN passes. However, the fraction of errors remaining is subject to being overestimated due to a decoder rarely ever giving corrections exactly the same as the errors that caused a given syndrome. To mitigate this effect, we instead compared the number of initial errors and number of errors after a given number of ANN passes with both sets of errors going through the heuristic error modification process. This removes trivial stabilizer loops and straightens out long chains to better estimate the number of errors which are effectively present in the the state before and after ANN decoding. Results are shown in Appendix Figure \ref{fig:errors_remaining}. We find that this gives plots quite similar to the fraction of stabilizers remaining. Simulations on systems with lower error rates, and a greater number of ANN passes have a lower average fraction of original errors remaining after ANN decoding. Average performance does not strongly change when investigating codes at distances 65 and 257. The fraction of errors remaining is often greater than the fraction of -1 stabilizers remaining. This can be understood as resulting from an effect that single residual strings of errors may be composed of several errors, but usual causes only two -1 stabilizers at the end points.
\noindent
\\ \\
\noindent
The error suppression properties of the surface code with the ANN+HDRG decoder can be investigated by plotting the error rate as a function of code distance. Such plots should show logical error rates able to be suppressed to arbitrarily low levels when larger codes are used. This is the case for the error rates investigated and shown in Appendix Figure \ref{fig:logical_vs_d}. Dotted lines are fits of curves of the form $p_L=ab^{(d-1)/2}$, where $p_L$, $d$, $a$, $b$ are the logical error rate, distance, first fit parameter and second fit parameter respectively.
\noindent
\\ \\
\noindent
The hybrid ANN+mop-up decoder performance can be investigated with other decoders acting as the mop-up decoder. Significant reduction in decoding time can occur in regimes where the ANN execution time can be negligible compared to the mop-up decoder execution time and the mop-up decoder execution time decreases with sparser syndromes. This is the case when using the PyMatching decoder as a mop-up decoder in Appendix Figure \ref{fig:mwpm_mopup}.
\noindent
\\ \\
\noindent
The performance of the ANN+HDRG decoder at low error rates and small code distances can be investigated by expanding the logical error rate to show the contributions from different weight errors. The expansion can be written,
\begin{equation}
    \mathrm{Pr}(\mathrm{Logical~error})=\sum_{i=0}^{n}\mathrm{Pr}({w}=i)\mathrm{Pr}(\mathrm{Logical~error}|w=i),
\end{equation}
where $\mathrm{Pr}(\mathrm{Logical~error})$ is the probability of having a logical error, $\mathrm{Pr}({w}=i)$ is the probability of the error weight being equal to $i$ and $\mathrm{Pr}(\mathrm{Logical~error}|w=i)$ is the conditional probability of a logical error occurring when the error rate is equal to $i$.  $\mathrm{Pr}({w}=i)$ can be calculated from the properties of the error model used, and is the probability mass function of the binomial distribution for the depolarizing noise model. $\mathrm{Pr}(\mathrm{Logical~error}|w=i)$ can be calculated from simulation. At low enough physical error rates, the logical error rates are dominated by the ability of a decoder to decode errors of weight less than or equal to $\mathrm{floor}((d-1)/2)$. These are the correctable errors of a code and decoders can be constructed to always decode these numbers of errors without causing a logical error. The first uncorrectable error has weight $\mathrm{floor}((d+1)/2)$ and dominates logical error rates at low physical error rates. The HDRG decoder, and other decoders which apply some corrections based on local information, can sometimes cause logical errors when decoding syndromes from errors which are otherwise correctable. The extent to which this may effect logical error rates in practice is shown in Appendix Figure \ref{fig:binom_error_rates}. The left hand side show binomial distributions for the number of errors which may occur under depolarizing noise with probability $0.1$ for codes of distance 3, 5 and 7. The parts of the distributions which correspond to correctable errors are highlighted with green rectangles. The conditional probabilities of causing a logical error during decoding is plotted for different numbers of qubits in error. An interesting observation is that these conditional error probabilities appear very similar to one another for different code distances when viewed with appropriately scaled x axes. On the right hand side is a plot of logical error rates for code of distance 3, 5, 7 and 9 as a function of physical error rates, when calculated with conditional error probabilities. Dashed lines show modified plots of logical error rates which would be achieved if correctable errors never caused logical errors when using the ANN+HDRG decoder. It can be observed that at lower physical error rates, the solid and dashed lines begin to deviate due to to the ANN+HDRG having a nonzero probability of incorrectly decoding correctable errors. The physical error rate when this begins to make a significant difference appears to decrease with increased code distance.
\noindent

\noindent
\section{Hardware and Software Details}\label{Appendix Hardware and Software Details}
\noindent
Performance benchmarks were evaluated on the NCI Gadi supercomputer. Jobs were submitted on the gpuvolta queue which uses Nvidia Tesla Volta V100-SXM2-32GB GPUs and  Intel Xeon Platinum 8268 (Cascade Lake) 2.9 GHz CPUs. Decoding latency measurements were performed with jobs scripts  using 12 CPU cores and 1 GPU.
\noindent
\\ \\
\noindent
Training and testing data was generated with the Python (version 3.9.2) and NumPy (version 1.20.0). HDRG decoding was performed in Python and also included the use of SciPy (version 1.6.2). 
\noindent
\\ \\
\noindent
ANN training and testing was performed using the Python Tensorflow library (version 2.6.0).
\noindent
\\ \\
\noindent
For MWPM decoding the Python PyMatching library (version 0.2.5) was used.

% \clearpage
% \newpage
%\bibliographystyle{unsrt}
% \def\bibsection{\subsection*{\refname}}
% \bibliographystyle{quantum}
% \bibliography{sample_sup}
%\clearpage

\section{Further Figures}\label{Appendix Further Figures}
\center
\begin{table}
\small
\caption{A comprehensive literature survey and the comparison of the machine learning based syndrome decoders.}\label{review_table}
\begin{tabular}{ |p{3cm}||p{2cm}|p{0.75cm}|p{1.8cm}|p{3.7 cm}|p{2.9cm}| }
 \hline
 \textbf{Paper}& \textbf{QEC Code} & \textbf{d$_{max}$} &\textbf{Threshold} & \textbf{ML Technique} & \textbf{Noise model }\\
 \hline
 Gicev, Hollenberg and Usman \textbf{[This Work]}   & Surface code with general boundaries  & 1025 &  0.138 & Supervised learning convolution neural network &  Depolarizing, inhomogeneous and biased noise \\
 \hline
 Meinerz, Park and Trebst \cite{Meinerz2021arxiv} &  Toric code & 255   &0.162(5) &  Supervised learning dense neural network & Depolarizing noise\\
 \hline
 Sweke \textit{et al.} \cite{sweke2020reinforcement} &Rotated surface code & 5 &Not reported & Reinforcement learning, Convolutional neural network & Bit-flip, Depolarizing, Phenomenological noise\\
 \hline
 Bhoumik \textit{et al.} \cite{bhoumik2021efficient} &  Rotated surface code & 7   & Not reported &  Supervised learning dense neural network & Multi-step depolarizing noise and biased noise.\\
 \hline
 Matekole \textit{et al.} \cite{matekole2022decoding} &  Toric code & 5   & Not reported &  Deep reinforcement learning & Bit flip\\
 \hline
 Overwater, Babaie and Sebastiano \cite{overwater2022neural} &  Rotated surface code & 9   & Not reported &  Supervised learning dense neural network & Depolarizing noise\\
 \hline
 Colomer, Skotiniotis and Muñoz-Tapia \cite{colomer2020reinforcement} &Toric code  & 9 &  0.103 &  Reinforcement learning, Deep convolutional net & Bit-flip \\
 \hline
 Ni \cite{ni2020neural} & Toric code  & 64 &  0.095 &  Supervised learning, Renormalization group based neural network & Bit flip \\
 \hline
 Sheth, Jafarzadeh and Gheorghiu \cite{sheth2020neural} &   Surface code  & 11 & Not reported & Supervised learning, Neural network ensemble learning & Depolarizing noise\\
 \hline
 Fitzek \textit{et al.} \cite{fitzek2020deep} &   Toric code  &9 &0.165 & Deep reinforcement learning & Depolarizing noise\\
 \hline
 Varsamopoulos, Bertels and Almudever \cite{varsamopoulos2020decoding} & Rotated surface code  & 9   &0.146 (depol.), 0.0032 (circ.)& Supervised learning, Feed forward neural networks, Recurrent neural nets with LSTMs & Depolarizing and circuit noise\\
 \hline
 Wagner, Kampermann and Bruß \cite{wagner2020symmetries} & Toric code  & 7 &Not reported&Supervised learning, Feed forward neural net &Depolarizing noise\\
 \hline
 Andreasson \textit{et al.} \cite{andreasson2019quantum} & Toric code  & 7 &Not reported & Deep reinforcement learning & Bit-flip\\
 \hline
 Varsamopoulos, Bertels and Almudever \cite{varsamopoulos2019comparing}  & Rotated surface code  & 9 &Not reported & Supervised learning, Neural network & Depolarizing noise\\
 \hline
 Chamberland and Ronagh \cite{chamberland2018deep} & Rotated surface code  & 5 &Not reported & Supervised learning, Deep neural networks, Single layer neural networks & Circuit noise\\
 \hline
 Breuckmann and Ni \cite{breuckmann2018scalable} & 3D toric code, 4D toric code  & 12& 0.175 (3D), 0.071 (4D) & Supervised learning, Convolutional neural network & Bit-flip, Phenomenological noise\\
 \hline
 Baireuther \textit{et al.} \cite{baireuther2018machine} & Rotated surface code  & 3&Not reported &  Supervised learning, Recurrent neural net with LSTMs & Depolarizing noise and Measurement errors\\
 \hline
 Varsamopoulos, Criger and Bertels \textit{et al.} \cite{varsamopoulos2017decoding} & Rotated surface code  & 7&Not reported & Supervised learning, Feed forward neural network & Bit-flip, Depolarizing, Phenomenological and Circuit noise\\
 \hline
 Krastanov and Jiang \cite{krastanov2017deep} & Toric code  & 9&0.164 &Neural net with 15-18 hidden layers  & Depolarizing noise\\
 \hline
 Torlai and Melko \cite{torlai2017neural} & Toric code  & 6&0.109 & Restricted Boltzmann machine & Phase-flip errors\\
 \hline
\end{tabular}

\end{table}

\clearpage

\clearpage

\begin{figure*}[htp]
\begin{center}
\resizebox{130mm}{!}{\includegraphics{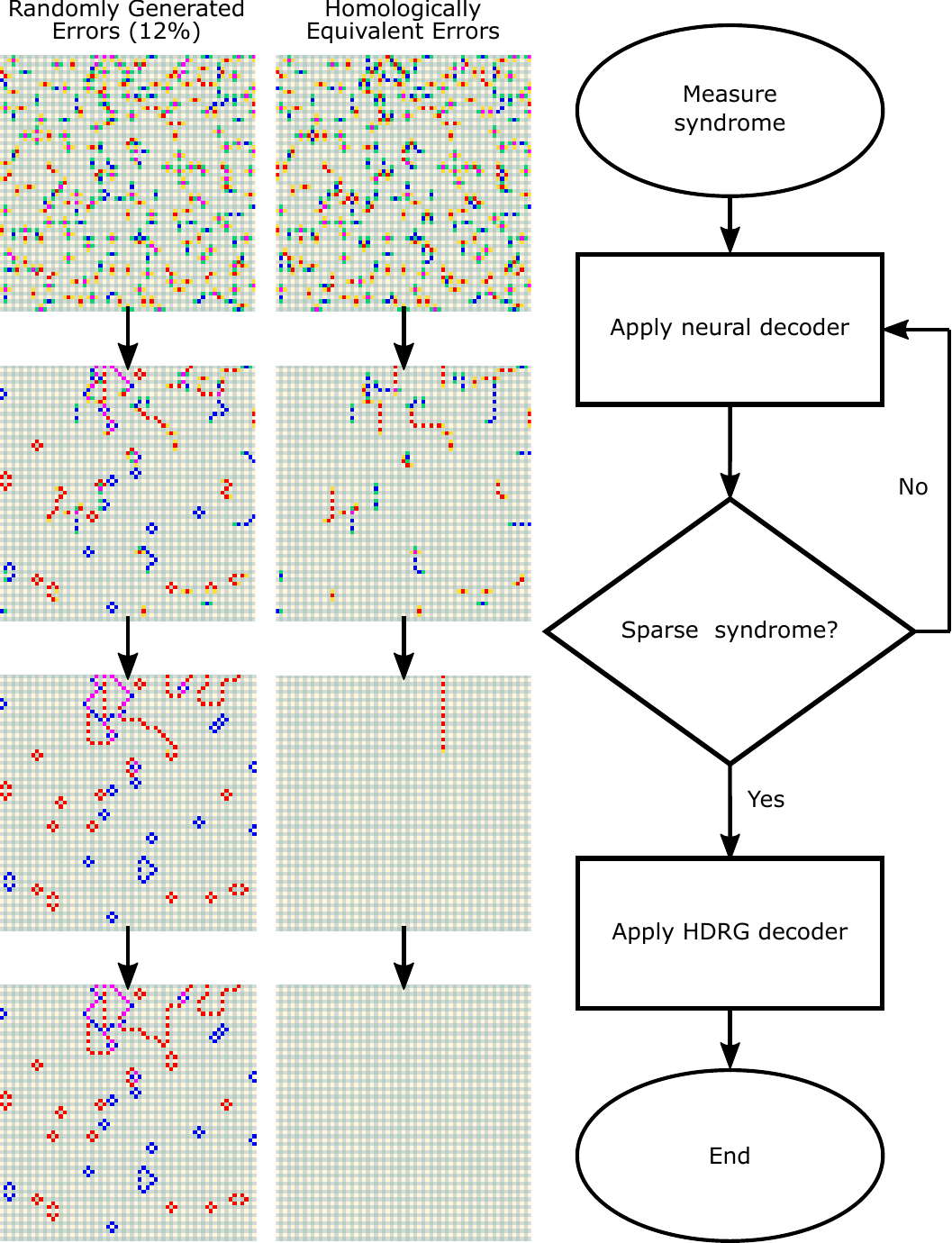}}
\end{center}
\caption{The operation of the full decoding procedure. Displayed on the left is an example of this procedure on a distance 33 surface code. First, the error syndrome is measured and given to the ANN decoder. The corrections are applied and the residual syndrome is given again to the ANN decoder until it becomes sparse (less than 10\% leftover errors). Once sparse, any leftover stabilizers are decoded by the HDRG decoder. Remaining errors, when occurring in loops are logically equivalent to an identity operation to states originally in the code space.  
}\label{fig:just_flowchart}
\end{figure*}

\begin{figure*}[htp]
\begin{center}
\resizebox{120mm}{!}{\includegraphics{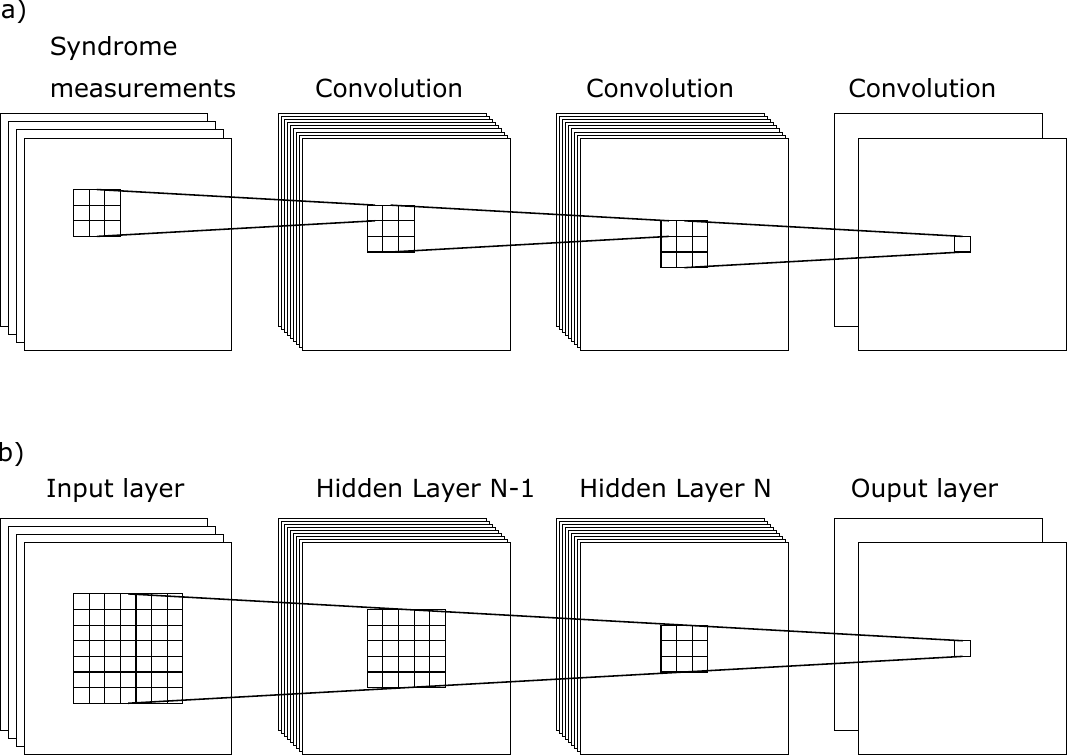}}
\end{center}
\caption{
    (a) The design of the convolutional neural network decoder presented. The input is understood as a 4 channel image, with the first two channels containing X and Z syndrome information and the last two channels containing X and Z boundary information. Two dimensional convolutional layers are use to form two hidden layers. The output layer is also a convolutional layer with the two output channels interpreted as the probability of an X or Z correction being appropriate at each qubit location. (b) Causal connectivity in neural networks with multiple adjacent convolutional layers. The output layer depends only on the neuron within a region resembling a light cone. This shows the effective window of local observation, or local receptive field, used to calculate which correction operator is appropriate at a particular physical qubit. 
}\label{fig:decoder_layers}
\end{figure*}

\begin{figure*}[htp]
\begin{center}
\resizebox{175mm}{!}{\includegraphics{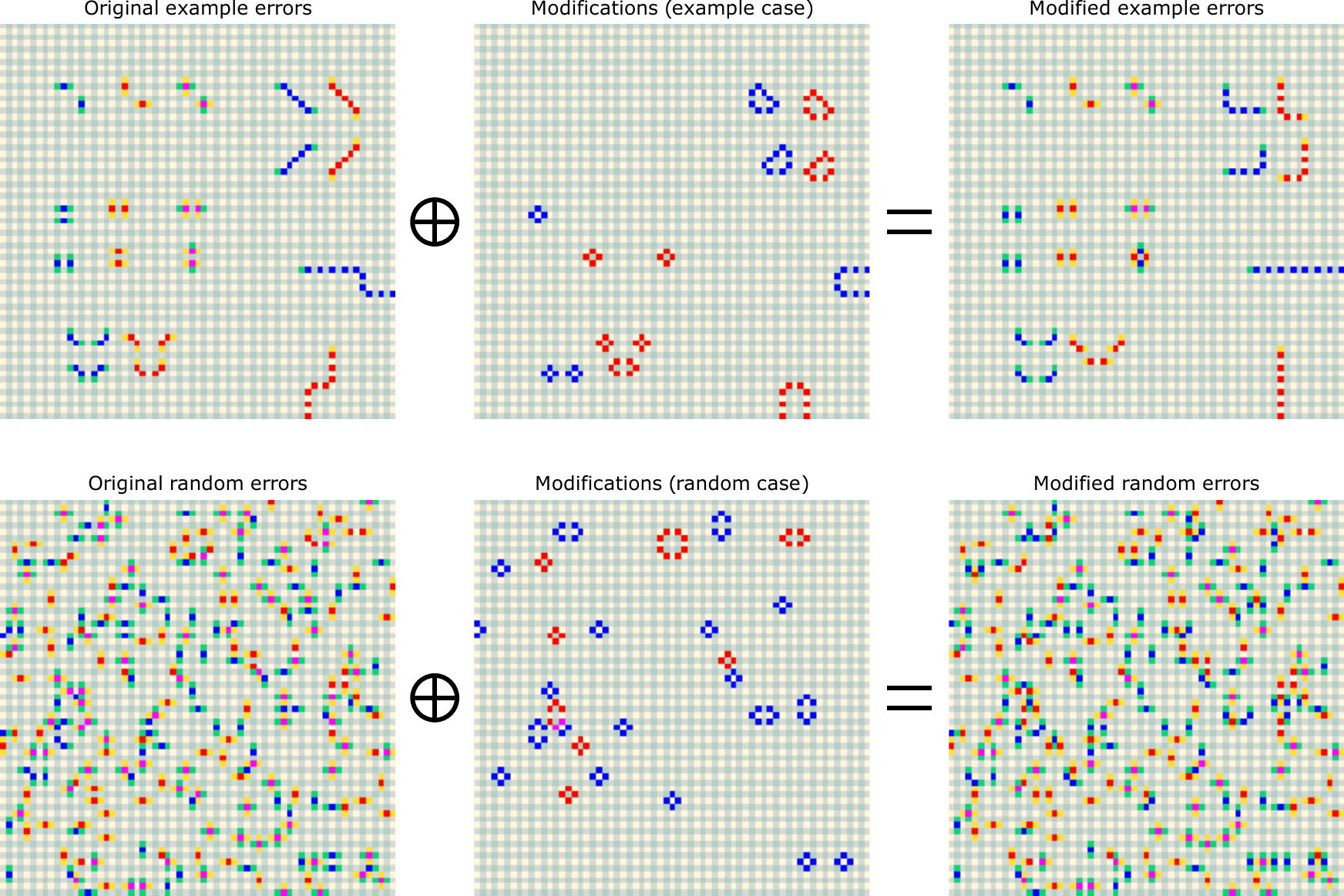}}
\end{center}
\caption{
    The heuristic error modification process. The top row shows how isolated errors, small clusters of errors and large chains of errors change under the error modification process. The bottom row shows this process applied to randomly generated errors.
}\label{fig:target_example}
\end{figure*}

\begin{figure*}[htp]
\begin{center}
\resizebox{175mm}{!}{\includegraphics{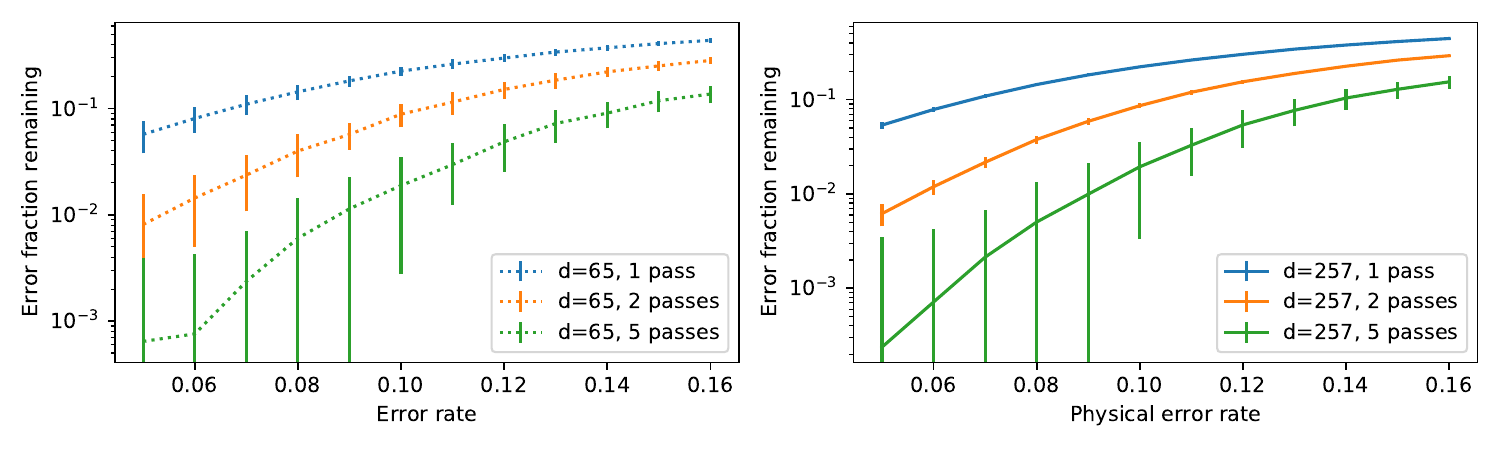}}
\end{center}
\caption{
    The fraction of errors remaining after ANN decoding and heuristic error modification. Data is shown for square surface codes of distance 65 and 257 suffering depolarizing noise. Error bars are 1 standard deviation in the fraction of errors remaining.
}\label{fig:errors_remaining}
\end{figure*}

\begin{figure*}[htp]
\begin{center}
\resizebox{155mm}{!}{\includegraphics{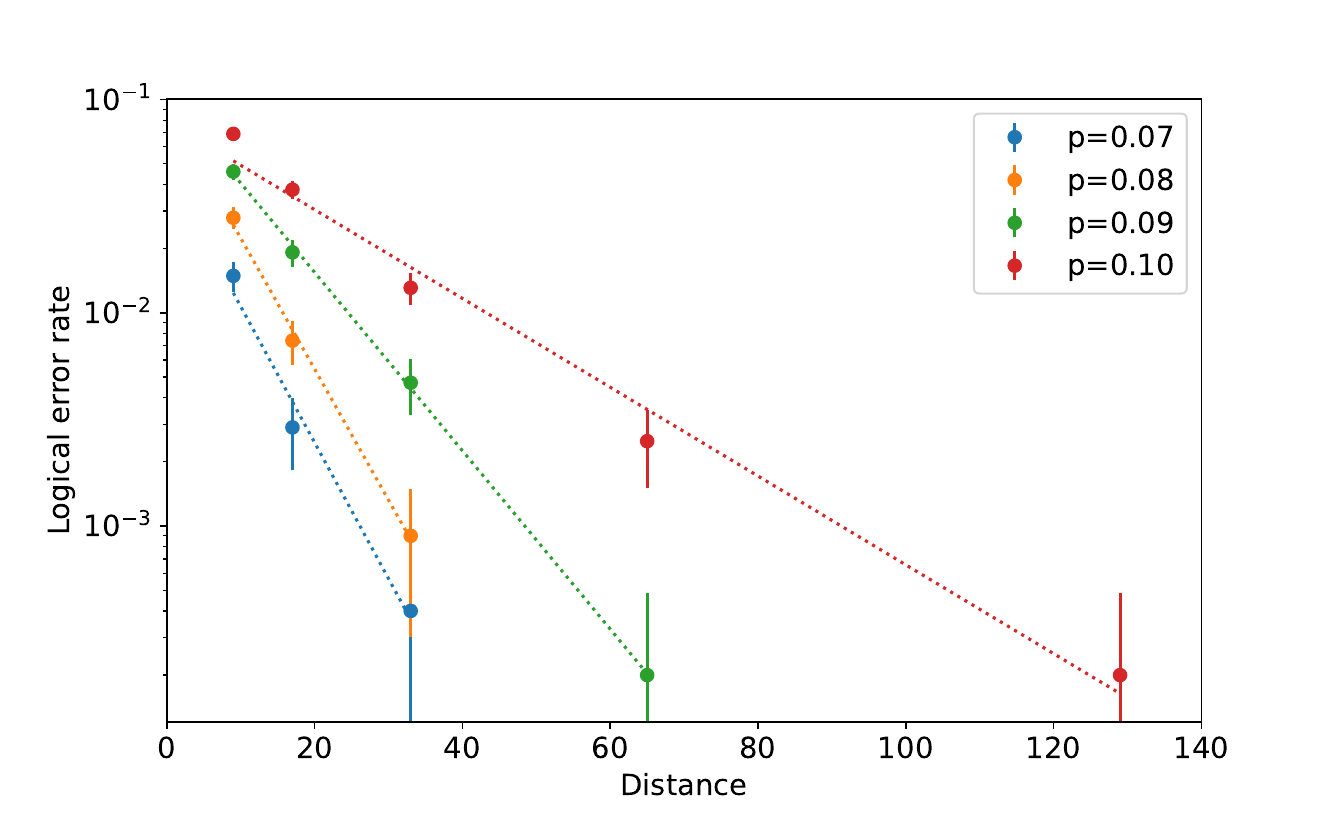}}
\end{center}
\caption{
    Logical error rate as a function of distance for surface codes suffering depolarizing noise when decoded by the 5$\times$ANN+HDRG decoder. Error bars correspond to a 2 standard deviation binomial confidence interval. Dotted lines are fits to exponential curves.
}\label{fig:logical_vs_d}
\end{figure*}

\begin{figure*}[htp]
\begin{center}
\resizebox{155mm}{!}{\includegraphics{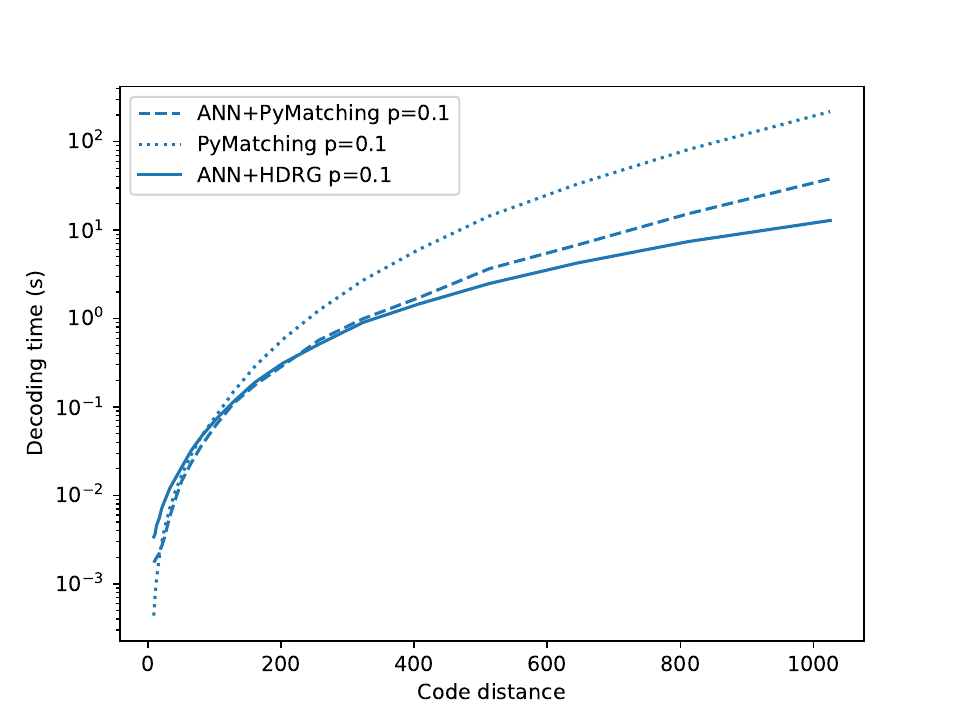}}
\end{center}
\caption{
    Decoding time as a function of code distance for PyMatching, ANN+PyMatching and ANN+HDRG. Results are show for the depolarizing noise model with an error rate of 0.1.
}\label{fig:mwpm_mopup}
\end{figure*}

\begin{figure*}[htp]
\begin{center}
\resizebox{175mm}{!}{\includegraphics{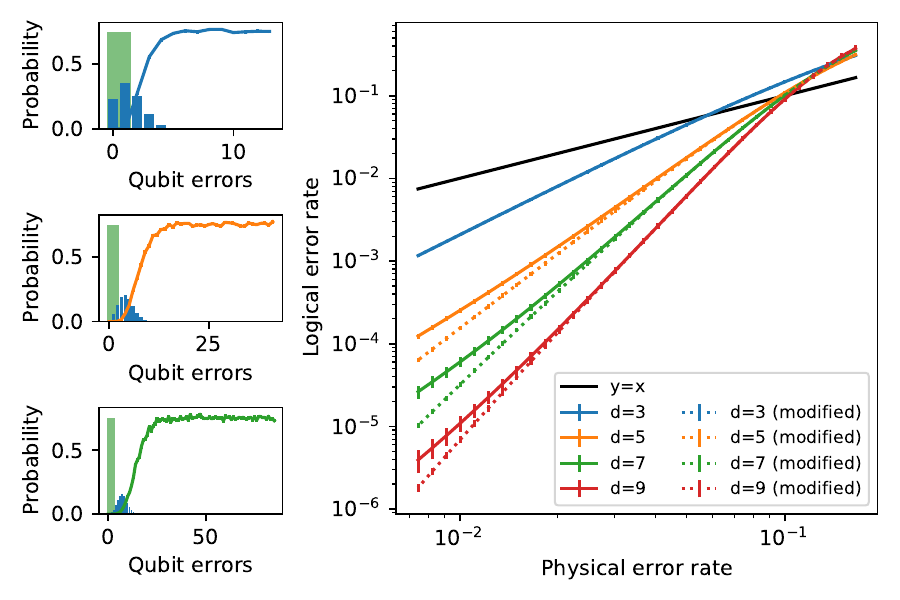}}
\end{center}
\caption{
    Low error rate performance for the ANN+HDRG decoder. Left hand side graphs show binomail distributions for the number of errors that occur for distance 3, 5 and 7 square surface codes experiencing errors with probability 0.1. Highlighted in green are numbers of errors below floor((d-1)/2). Curves are plotted showing the conditional probability of an error occuring when a given number of qubits are in error. The right hand side shows logical error rate as a function of physical error rate when calculated with conditional probabilities. Dashed lines show logical error rates if the conditional probabilities of error when floor((d-1)/2) or fewer qubit errors occur is zero. Error bars correspond to 1 standard deviation binomial confidence intervals.
}\label{fig:binom_error_rates}
\end{figure*}

\begin{figure*}[htp]
\begin{center}
\resizebox{170mm}{!}{\includegraphics{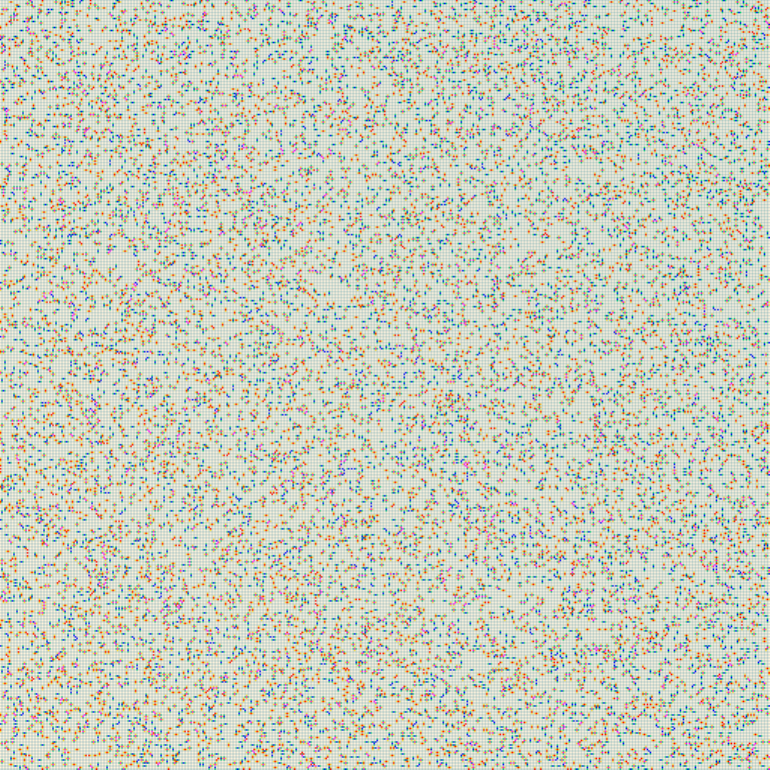}}
\end{center}
\caption{
   An example of  distance 257 surface code subject to 12\% depolarizing noise.
}\label{fig:257_12_initial}
\end{figure*}

\begin{figure*}[htp]
\begin{center}
\resizebox{170mm}{!}{\includegraphics{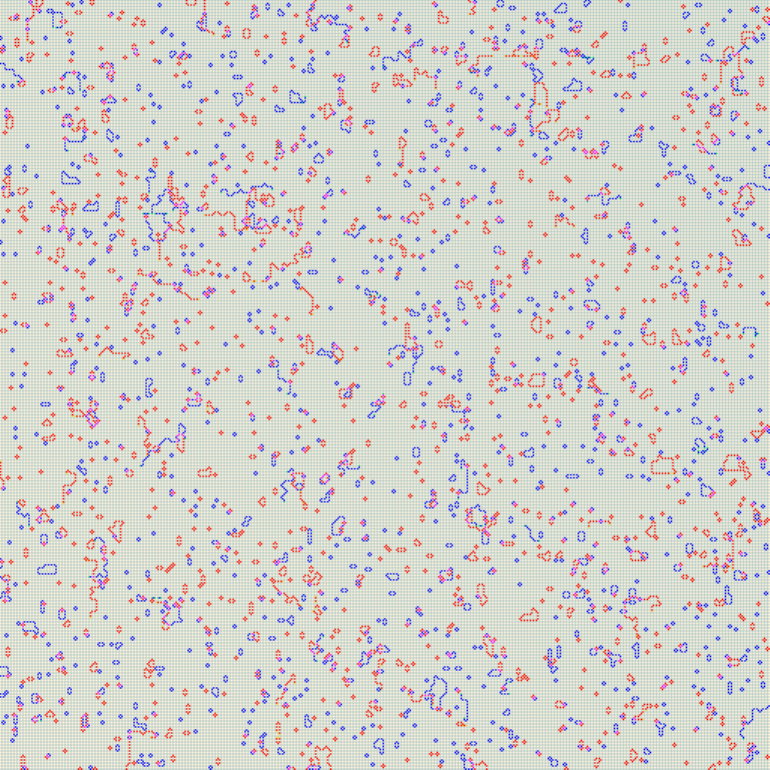}}
\end{center}
\caption{
    The distance 257 surface code subject to 12\% depolarizing noise (from Figure~\ref{fig:257_12_initial}) after five passes through the ANN decoder.
}\label{fig:257_12_5_pass}
\end{figure*}

\begin{figure*}[htp]
\begin{center}
\resizebox{170mm}{!}{\includegraphics{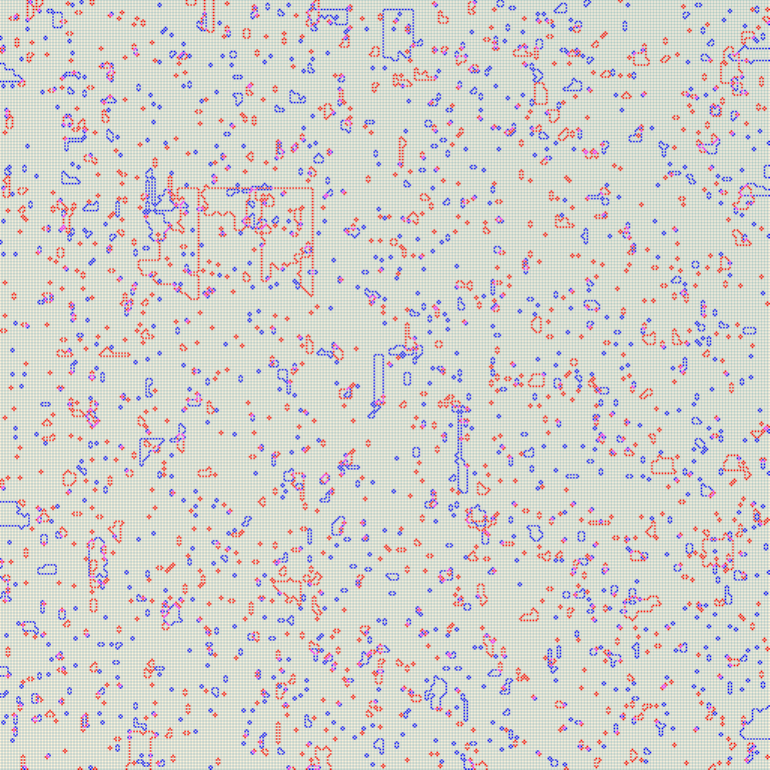}}
\end{center}
\caption{
The distance 257 surface code subject to 12\% depolarizing noise (from Figure~\ref{fig:257_12_initial}) after five passes through the ANN decoder, followed by one mop-up pass through the HDRG decoder.
}\label{fig:257_12_hdrg}
\end{figure*}

\begin{figure*}[htp]
\begin{center}
\resizebox{170mm}{!}{\includegraphics{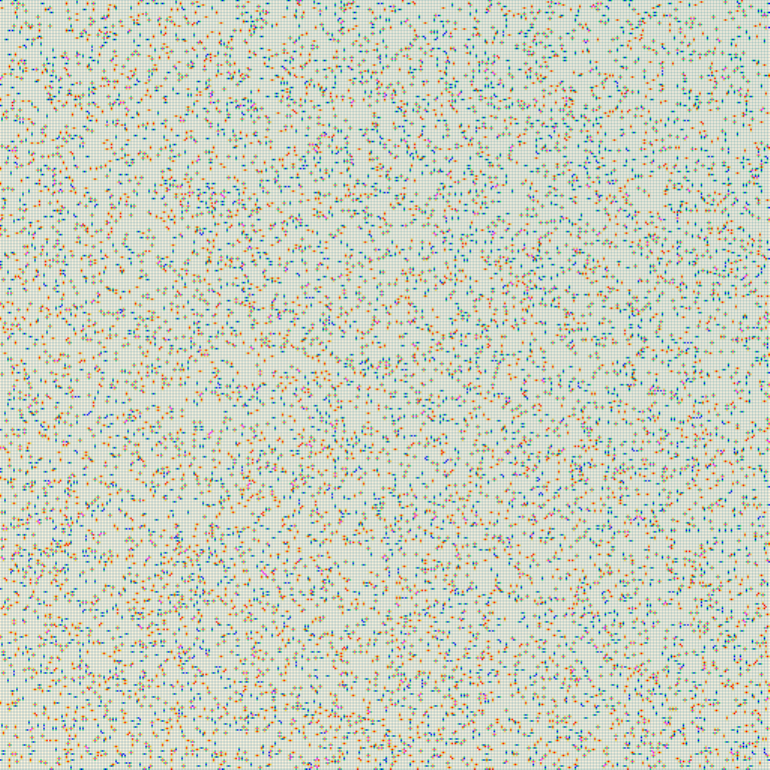}}
\end{center}
\caption{
    An example of  distance 257 surface code subject to 8\% depolarizing noise.
}\label{fig:257_08_initial}
\end{figure*}

\begin{figure*}[htp]
\begin{center}
\resizebox{170mm}{!}{\includegraphics{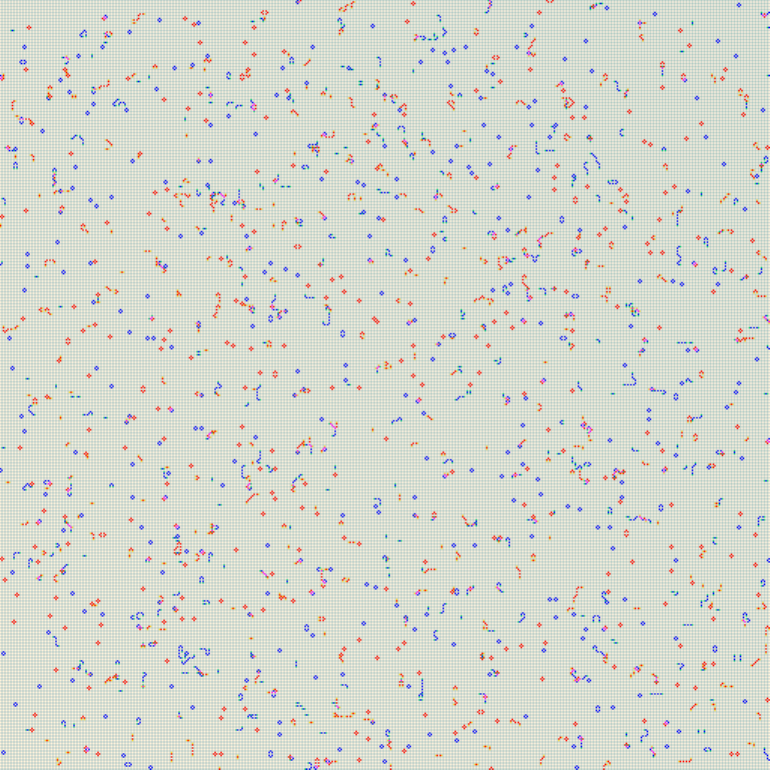}}
\end{center}
\caption{
    The distance 257 surface code subject to 8\% depolarizing noise (from Figure~\ref{fig:257_08_initial}) after a single pass through the ANN decoder.
}\label{fig:257_08_1_pass}
\end{figure*}

\begin{figure*}[htp]
\begin{center}
\resizebox{170mm}{!}{\includegraphics{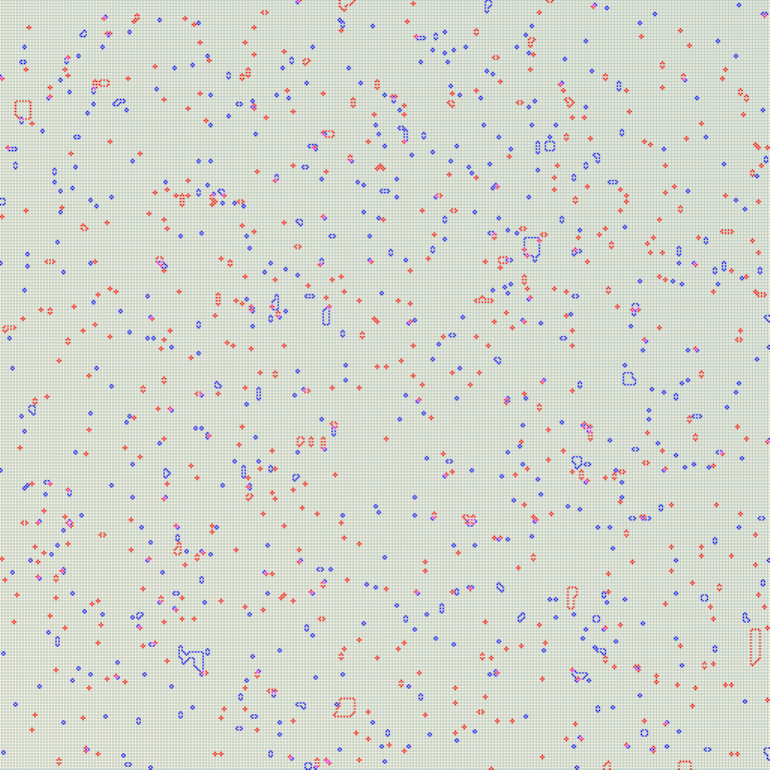}}
\end{center}
\caption{
    The distance 257 surface code subject to 8\% depolarizing noise (from Figure~\ref{fig:257_08_initial}) after a single pass through the ANN decoder, followed by mop-up by the HDRG decoder.
}\label{fig:257_08_hdrg}
\end{figure*}

\begin{figure*}[htp]
\begin{center}
\resizebox{170mm}{!}{\includegraphics{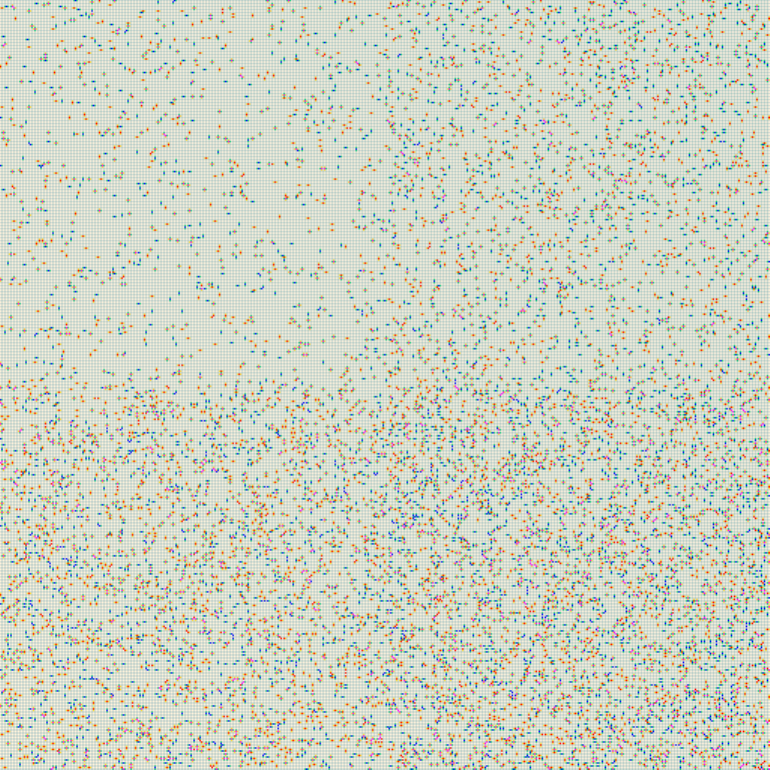}}
\end{center}
\caption{
    An example of  distance 257 surface code subject to spatially inhomogeneous noise where the board is divided into four patches with depolarising noise levels of 3\%, 6\%, 9\% and 12\%.
}\label{fig:inhom_2_2_flowchart1}
\end{figure*}

\begin{figure*}[htp]
\begin{center}
\resizebox{170mm}{!}{\includegraphics{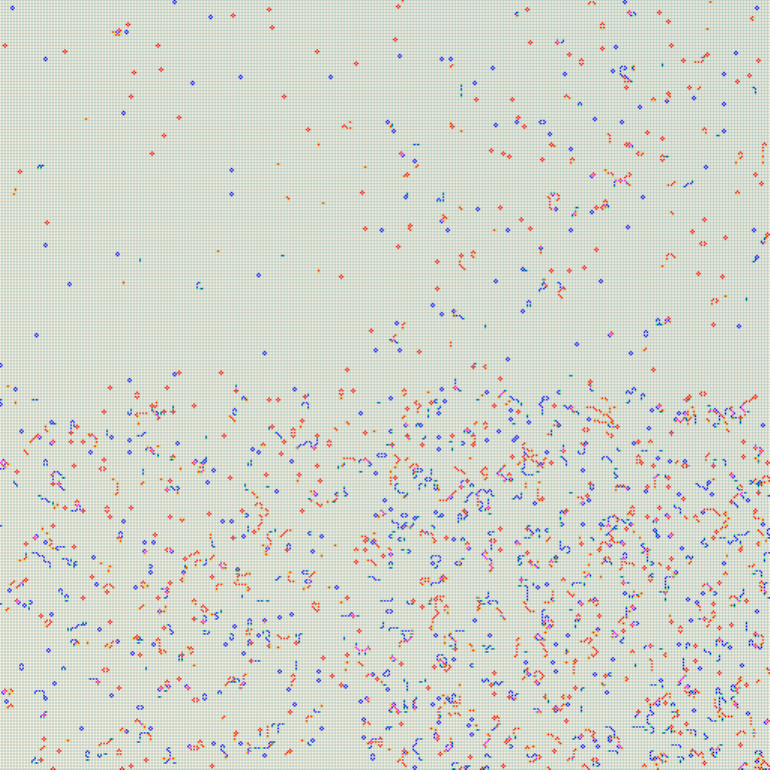}}
\end{center}
\caption{
    The distance 257 surface code subject to inhomogeneous noise (from Figure~\ref{fig:inhom_2_2_flowchart1}) after a single pass through the ANN decoder.
}\label{fig:inhom_2_2_flowchart2}
\end{figure*}

\begin{figure*}[htp]
\begin{center}
\resizebox{170mm}{!}{\includegraphics{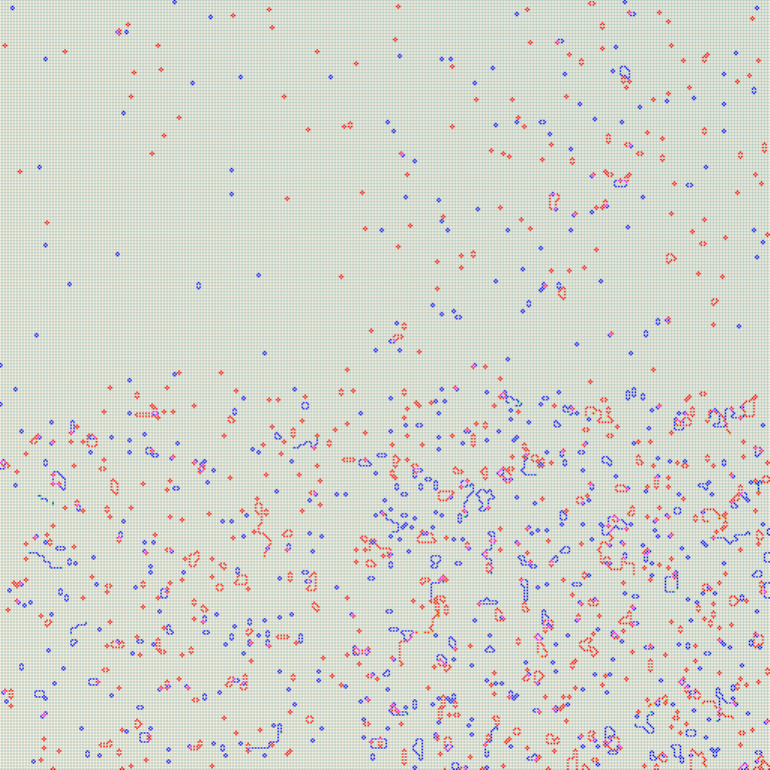}}
\end{center}
\caption{
    The distance 257 surface code subject to inhomogeneous noise (from Figure~\ref{fig:inhom_2_2_flowchart1}) after five passes through the ANN decoder.
}\label{fig:inhom_2_2_flowchart3}
\end{figure*}

\begin{figure*}[htp]
\begin{center}
\resizebox{170mm}{!}{\includegraphics{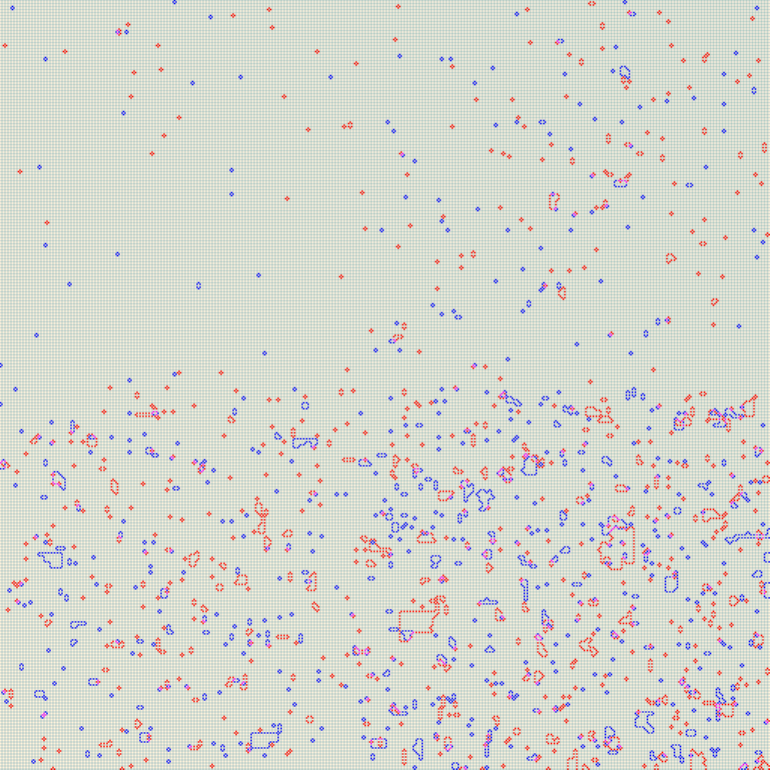}}
\end{center}
\caption{
    The distance 257 surface code subject to inhomogeneous noise (from Figure~\ref{fig:inhom_2_2_flowchart1}) after five passes through the ANN decoder, followed by mop-up by the HDRG decoder.
}\label{fig:inhom_2_2_flowchart4}
\end{figure*}

\clearpage

\begin{figure*}[htp]
\begin{center}
\resizebox{170mm}{!}{\includegraphics{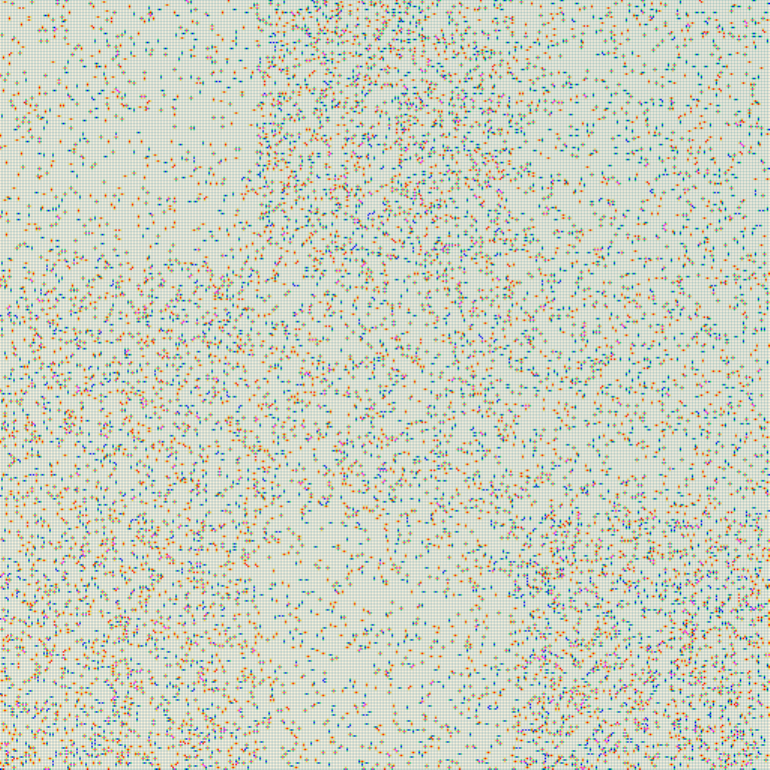}}
\end{center}
\caption{
    An example of  distance 257 surface code subject to spatially inhomogeneous noise where the board is divided into nine patches with depolarising noise levels of 4\%, 11\%, 6\%, 9\%, 8\%, 7\%, 1\%, 5\%  and 12\%.
}\label{fig:inhom_3_3_flowchart1}
\end{figure*}

\begin{figure*}[htp]
\begin{center}
\resizebox{170mm}{!}{\includegraphics{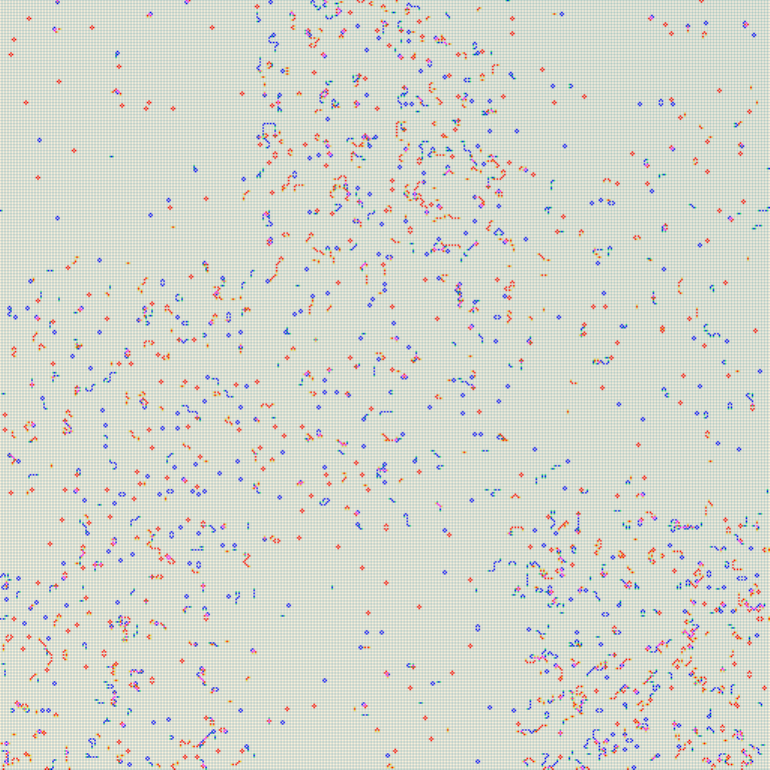}}
\end{center}
\caption{
    The distance 257 surface code subject to inhomogeneous noise (from Figure~\ref{fig:inhom_3_3_flowchart1}) after a single pass through the ANN decoder.
}\label{fig:inhom_3_3_flowchart2}
\end{figure*}

\begin{figure*}[htp]
\begin{center}
\resizebox{170mm}{!}{\includegraphics{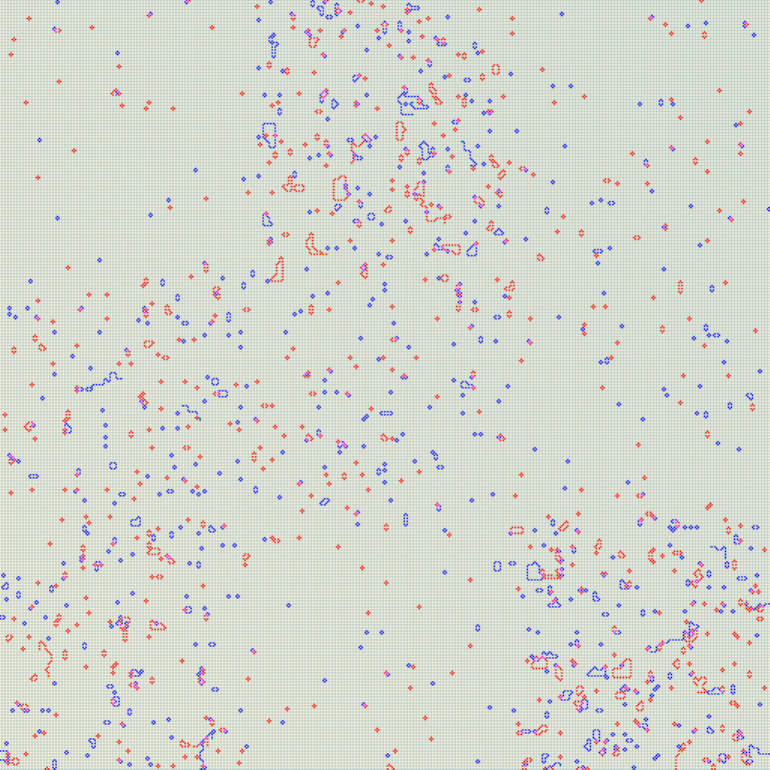}}
\end{center}
\caption{
    The distance 257 surface code subject to inhomogeneous noise (from Figure~\ref{fig:inhom_3_3_flowchart1}) after five passes through the ANN decoder.
}\label{fig:inhom_3_3_flowchart3}
\end{figure*}

\begin{figure*}[htp]
\begin{center}
\resizebox{170mm}{!}{\includegraphics{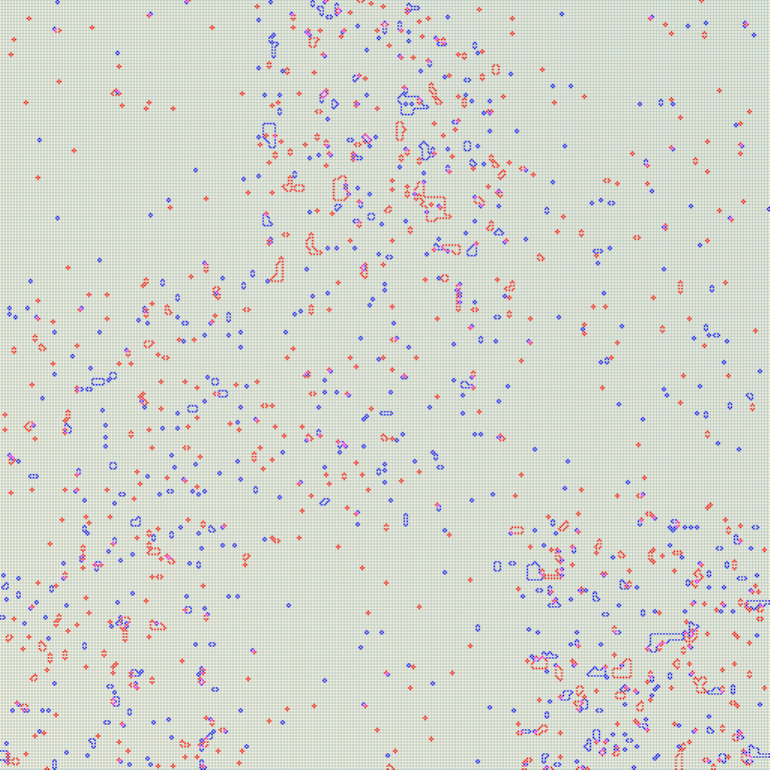}}
\end{center}
\caption{
    The distance 257 surface code subject to inhomogeneous noise (from Figure~\ref{fig:inhom_3_3_flowchart1}) after five passes through the ANN decoder, followed by mop-up by the HDRG decoder.
}\label{fig:inhom_3_3_flowchart4}
\end{figure*}

\clearpage

\begin{figure*}[htp]
\begin{center}
\resizebox{170mm}{!}{\includegraphics{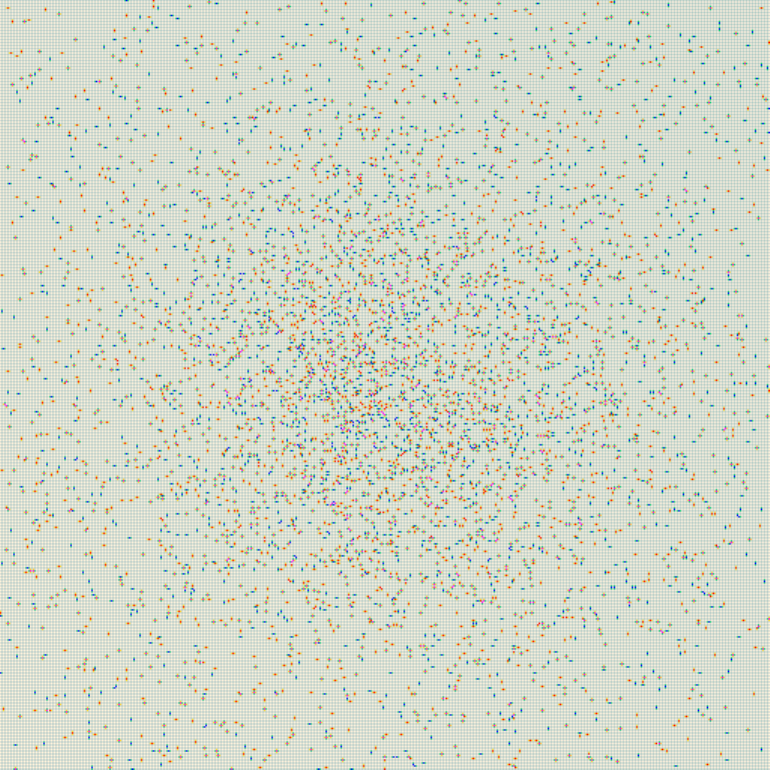}}
\end{center}
\caption{
    An example of  distance 257 surface code subject to spatially inhomogeneous noise where the error rate at the center of board is significantly higher than the error rates closer to the edges.
}\label{fig:inhom_radial_flowchart1}
\end{figure*}

\begin{figure*}[htp]
\begin{center}
\resizebox{170mm}{!}{\includegraphics{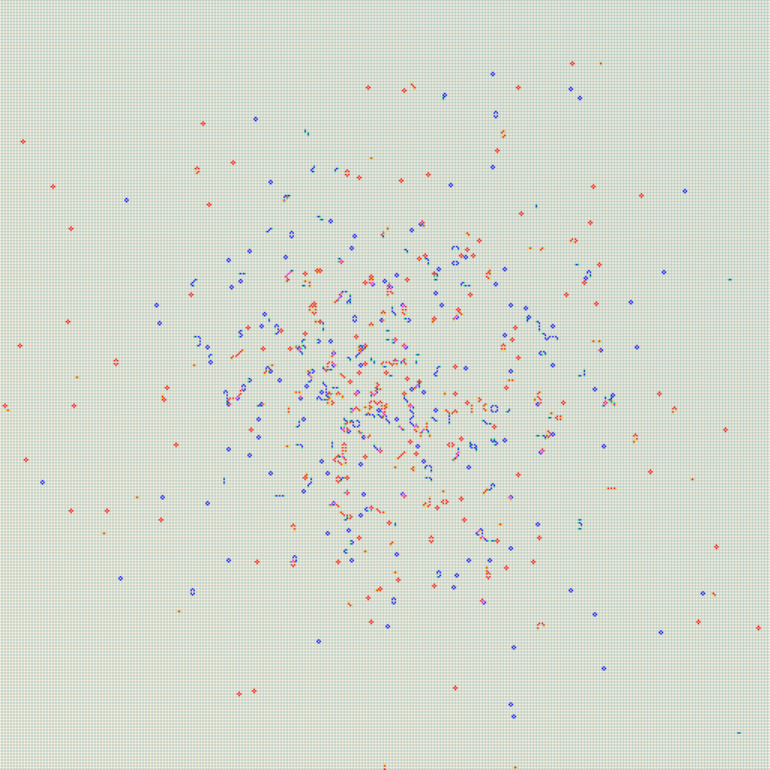}}
\end{center}
\caption{
    The distance 257 surface code subject to inhomogeneous noise (from Figure~\ref{fig:inhom_radial_flowchart1}) after a single pass through the ANN decoder.
}\label{fig:inhom_radial_flowchart2}
\end{figure*}

\begin{figure*}[htp]
\begin{center}
\resizebox{170mm}{!}{\includegraphics{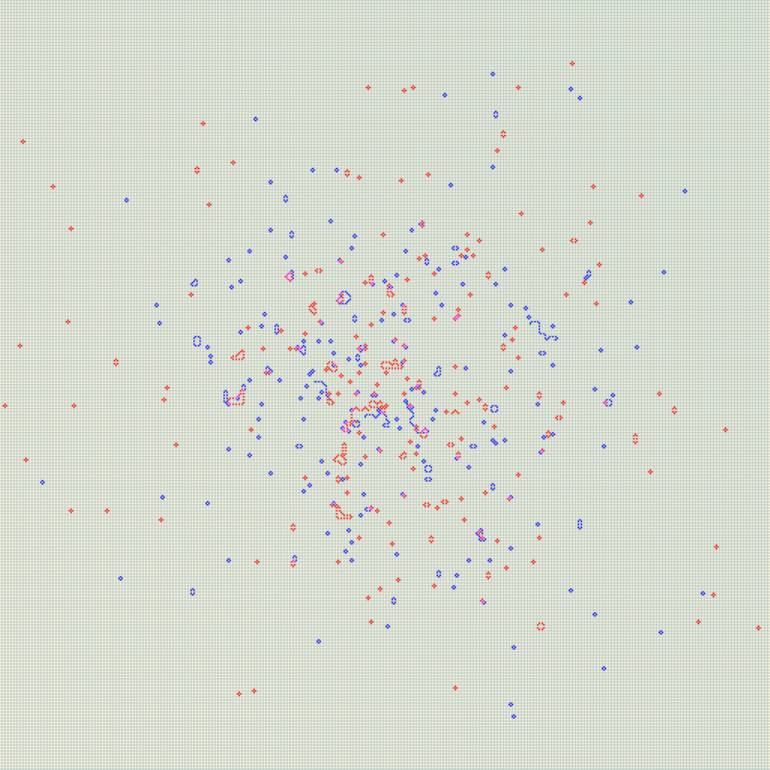}}
\end{center}
\caption{
    The distance 257 surface code subject to inhomogeneous noise (from Figure~\ref{fig:inhom_radial_flowchart1}) after five passes through the ANN decoder.
}\label{fig:inhom_radial_flowchart3}
\end{figure*}

\begin{figure*}[htp]
\begin{center}
\resizebox{170mm}{!}{\includegraphics{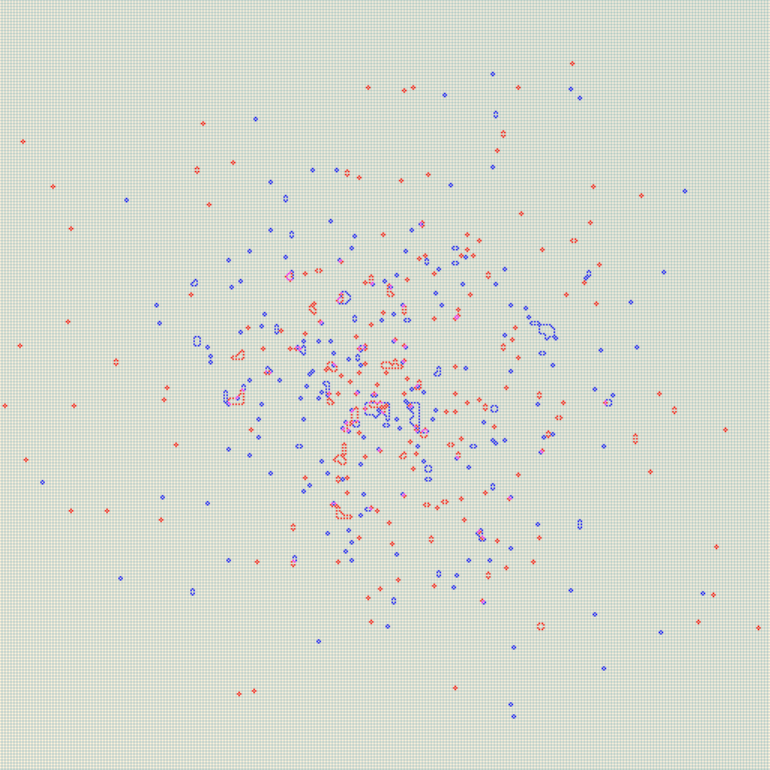}}
\end{center}
\caption{
    The distance 257 surface code subject to inhomogeneous noise (from Figure~\ref{fig:inhom_radial_flowchart1}) after five passes through the ANN decoder, followed by mop-up by the HDRG decoder.
}\label{fig:inhom_radial_flowchart4}
\end{figure*}

\begin{figure*}[htp]
\begin{center}
\resizebox{170mm}{!}{\includegraphics{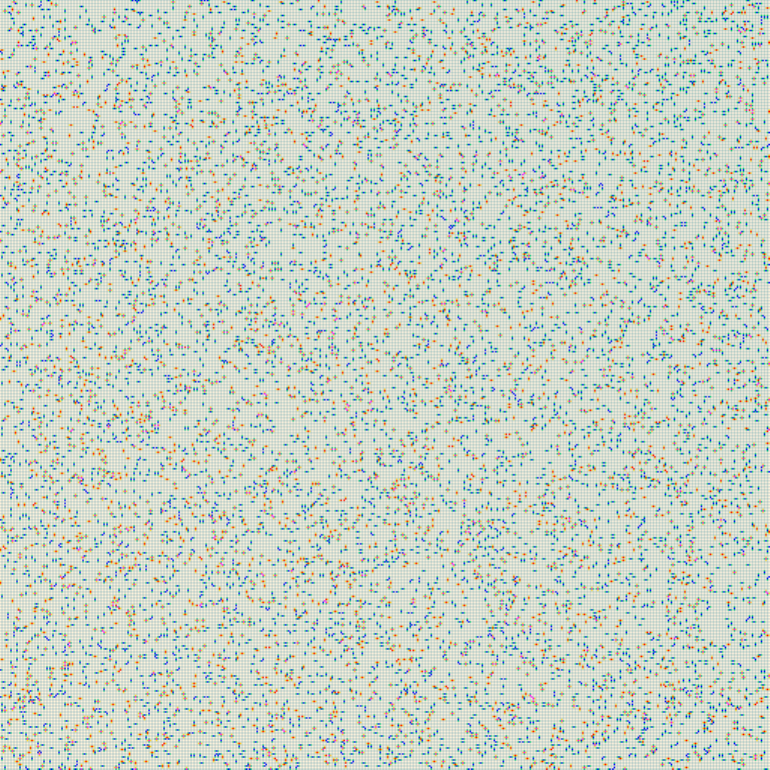}}
\end{center}
\caption{
    An example of distance 257 surface code subject to randomly distributed biased noise (8\%), where the error probability for X errors is 60\% compared to 20\% for Y and Z errors each.
}\label{fig:biased_60_20_20_flowchart1}
\end{figure*}

\begin{figure*}[htp]
\begin{center}
\resizebox{170mm}{!}{\includegraphics{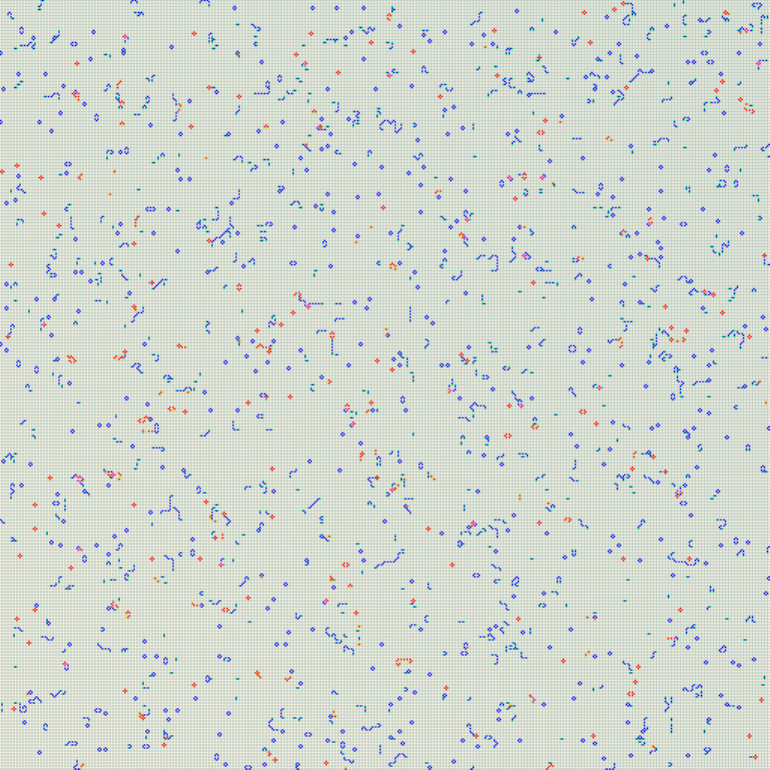}}
\end{center}
\caption{
    The distance 257 surface code subject to biased noise (from Figure~\ref{fig:biased_60_20_20_flowchart1}) after a single pass through the ANN decoder.
}\label{fig:biased_60_20_20_flowchart2}
\end{figure*}

\begin{figure*}[htp]
\begin{center}
\resizebox{170mm}{!}{\includegraphics{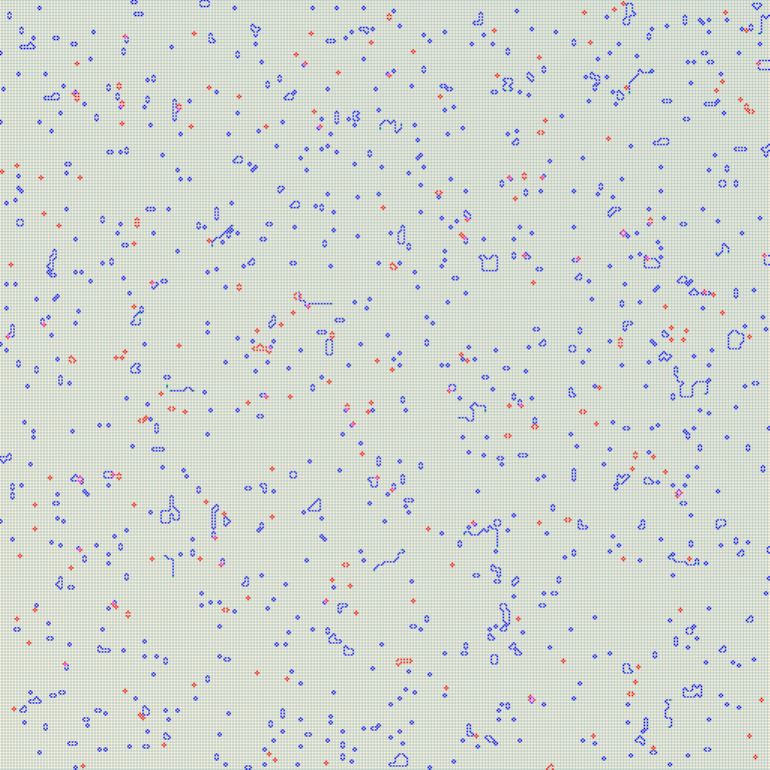}}
\end{center}
\caption{
    The distance 257 surface code subject to biased noise (from Figure~\ref{fig:biased_60_20_20_flowchart1}) after five passes through the ANN decoder.
}\label{fig:biased_60_20_20_flowchart3}
\end{figure*}

\begin{figure*}[htp]
\begin{center}
\resizebox{170mm}{!}{\includegraphics{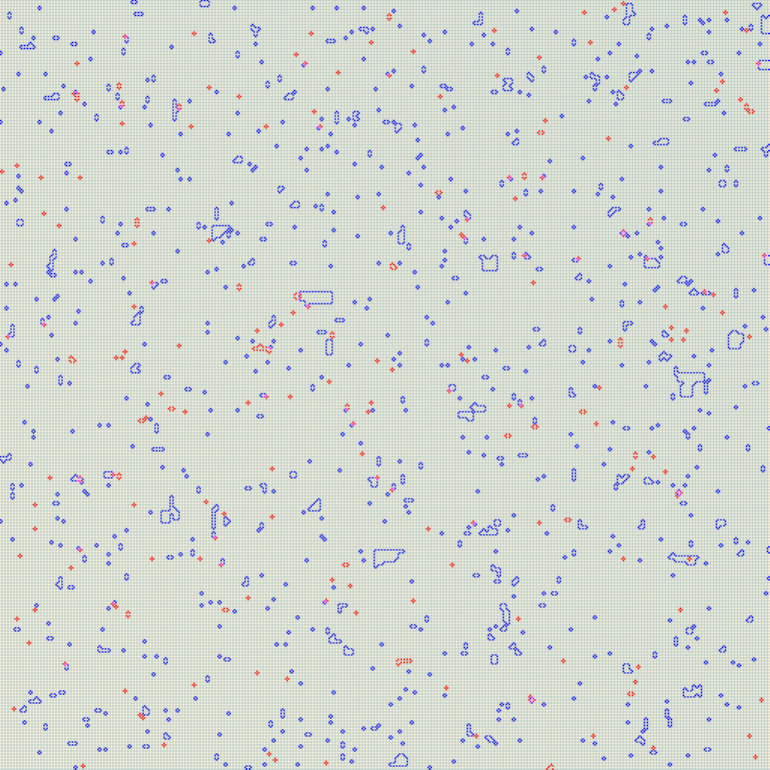}}
\end{center}
\caption{
    The distance 257 surface code subject to biased noise (from Figure~\ref{fig:biased_60_20_20_flowchart1}) after five passes through the ANN decoder, followed by mop-up by the HDRG decoder.
}\label{fig:biased_60_20_20_flowchart4}
\end{figure*}

\clearpage

\begin{figure*}[htp]
\begin{center}
\resizebox{170mm}{!}{\includegraphics{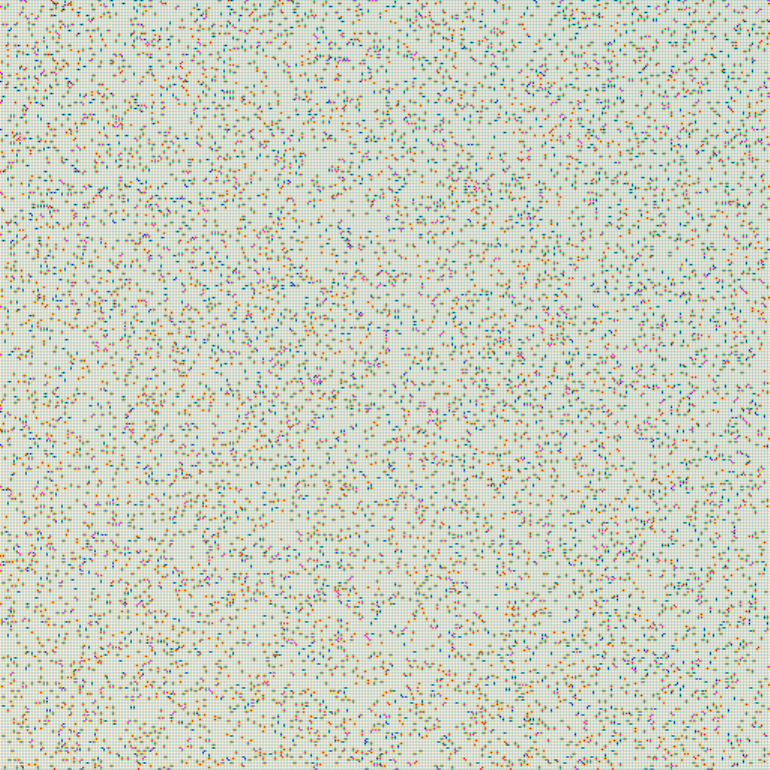}}
\end{center}
\caption{
    An example of distance 257 surface code subject to randomly distributed biased noise (8\%), where the error rate for Y errors is 60\% compared to 20\% for X and Z errors each.
}\label{fig:biased_20_60_20_flowchart1}
\end{figure*}

\begin{figure*}[htp]
\begin{center}
\resizebox{170mm}{!}{\includegraphics{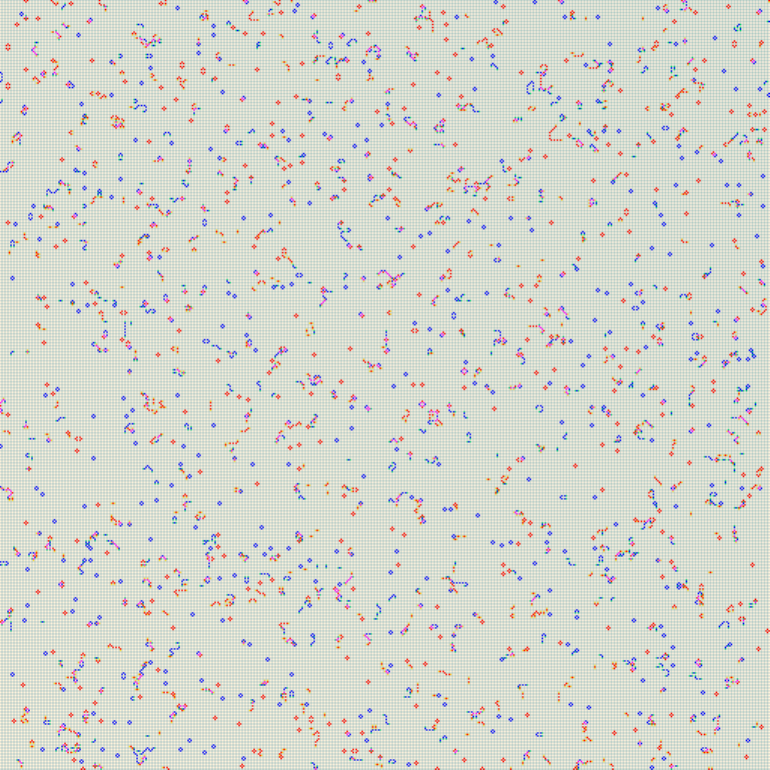}}
\end{center}
\caption{
    The distance 257 surface code subject to biased noise (from Figure~\ref{fig:biased_20_60_20_flowchart1}) after a single pass through the ANN decoder.
}\label{fig:biased_20_60_20_flowchart2}
\end{figure*}

\begin{figure*}[htp]
\begin{center}
\resizebox{170mm}{!}{\includegraphics{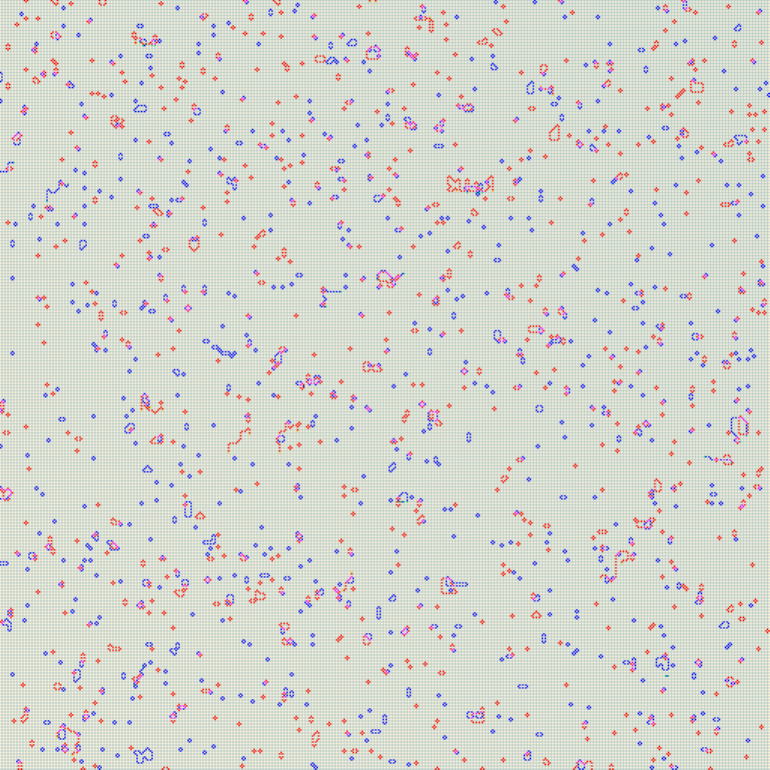}}
\end{center}
\caption{
    The distance 257 surface code subject to biased noise (from Figure~\ref{fig:biased_20_60_20_flowchart1}) after five passes through the ANN decoder.
}\label{fig:biased_20_60_20_flowchart3}
\end{figure*}

\begin{figure*}[htp]
\begin{center}
\resizebox{170mm}{!}{\includegraphics{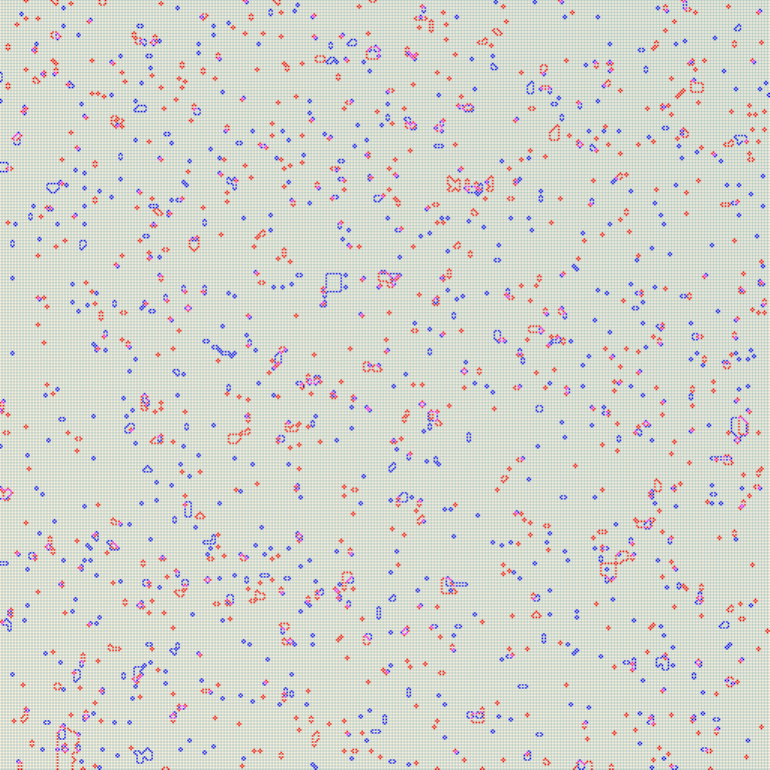}}
\end{center}
\caption{
    The distance 257 surface code subject to biased noise (from Figure~\ref{fig:biased_20_60_20_flowchart1}) after five passes through the ANN decoder, followed by mop-up by the HDRG decoder.
}\label{fig:biased_20_60_20_flowchart4}
\end{figure*}

\clearpage

\begin{figure*}[htp]
\begin{center}
\resizebox{170mm}{!}{\includegraphics{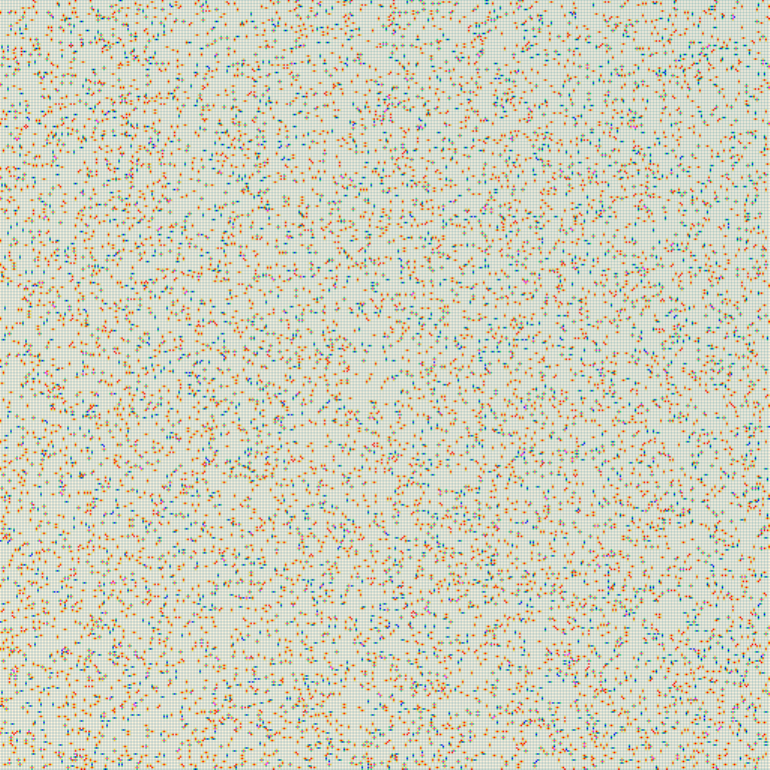}}
\end{center}
\caption{
    An example of distance 257 surface code subject to randomly distributed biased noise (8\%), where the error rate for Z errors is 60\% compared to 20\% for X and Y errors each.
}\label{fig:biased_20_20_60_flowchart1}
\end{figure*}

\begin{figure*}[htp]
\begin{center}
\resizebox{170mm}{!}{\includegraphics{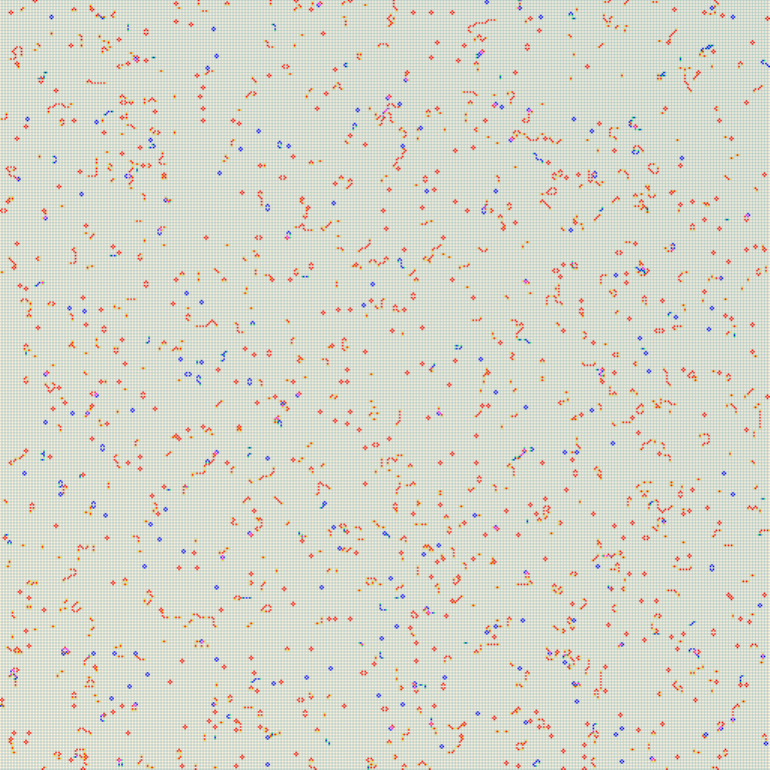}}
\end{center}
\caption{
    The distance 257 surface code subject to biased noise (from Figure~\ref{fig:biased_20_20_60_flowchart1}) after a single pass through the ANN decoder.
}\label{fig:biased_20_20_60_flowchart2}
\end{figure*}

\begin{figure*}[htp]
\begin{center}
\resizebox{170mm}{!}{\includegraphics{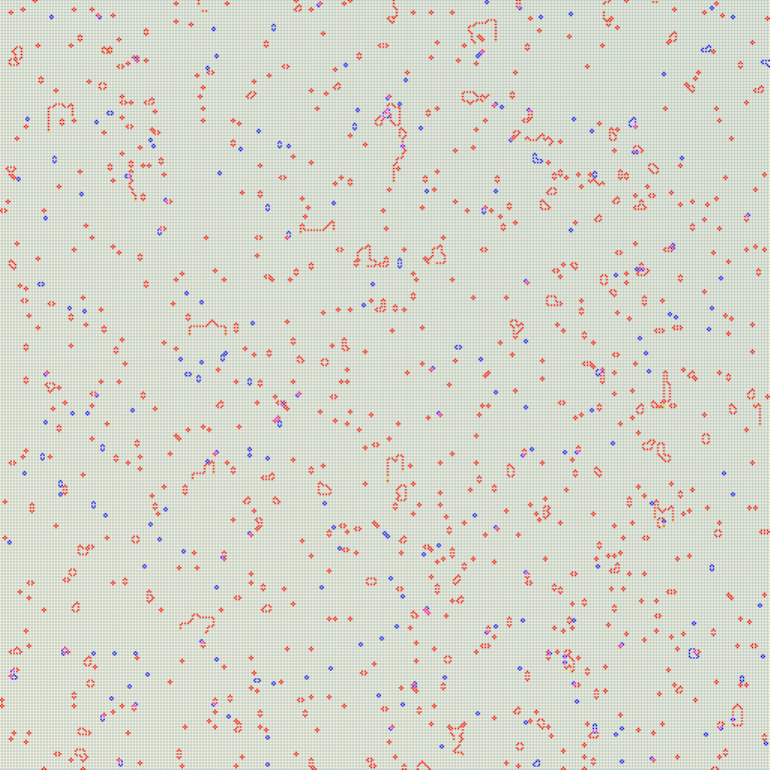}}
\end{center}
\caption{
    The distance 257 surface code subject to biased noise (from Figure~\ref{fig:biased_20_20_60_flowchart1}) after five passes through the ANN decoder.
}\label{fig:biased_20_20_60_flowchart3}
\end{figure*}

\begin{figure*}[htp]
\begin{center}
\resizebox{170mm}{!}{\includegraphics{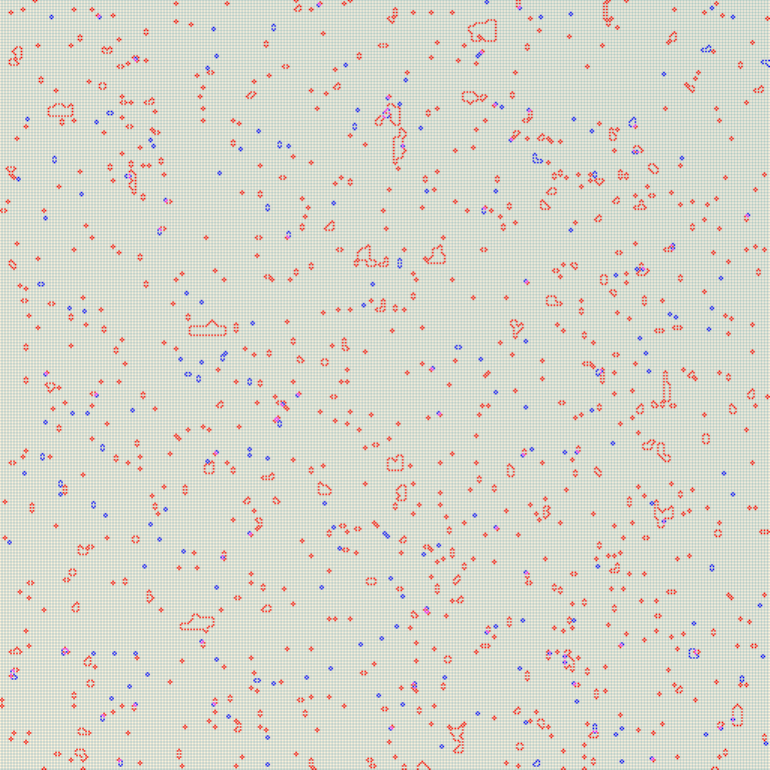}}
\end{center}
\caption{
    The distance 257 surface code subject to biased noise (from Figure~\ref{fig:biased_20_20_60_flowchart1}) after five passes through the ANN decoder, followed by mop-up by the HDRG decoder.
}\label{fig:biased_20_20_60_flowchart4}
\end{figure*}

\clearpage

\begin{figure*}[htp]
\begin{center}
\resizebox{170mm}{!}{\includegraphics{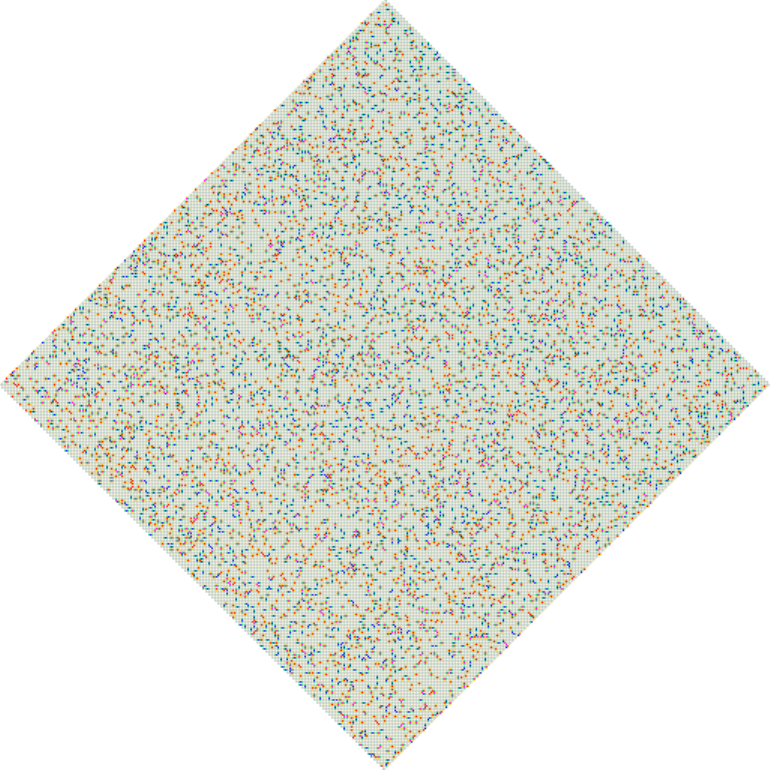}}
\end{center}
\caption{
    An example of distance 257 rotated surface code subject to randomly distributed depolarising noise (12\%).
}\label{fig:rot_12_flowchart1}
\end{figure*}

\begin{figure*}[htp]
\begin{center}
\resizebox{170mm}{!}{\includegraphics{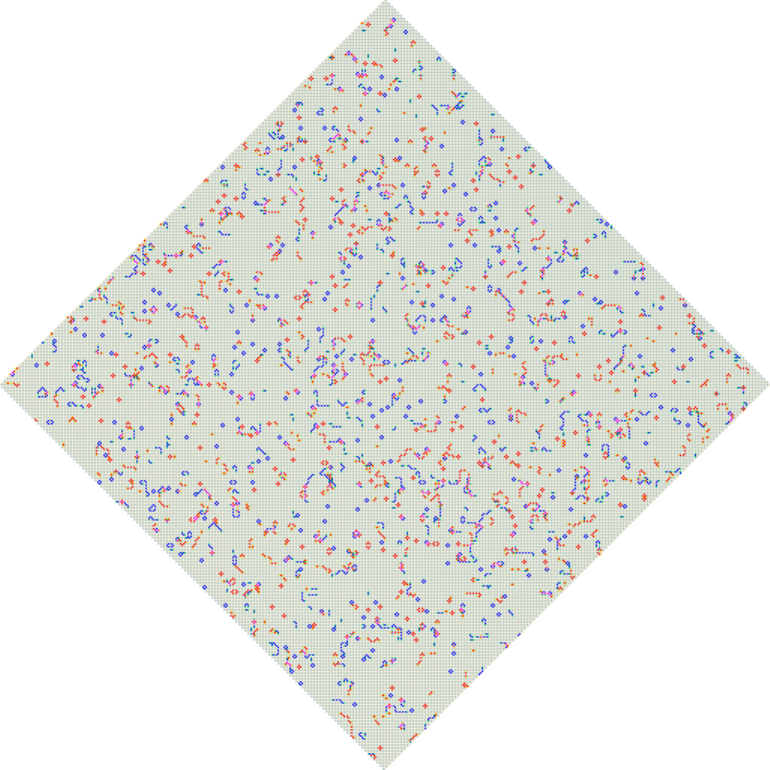}}
\end{center}
\caption{
    The distance 257 rotated surface code subject to depolarising noise (from Figure~\ref{fig:rot_12_flowchart2}) after a single pass through the ANN decoder.
}\label{fig:rot_12_flowchart2}
\end{figure*}

\begin{figure*}[htp]
\begin{center}
\resizebox{170mm}{!}{\includegraphics{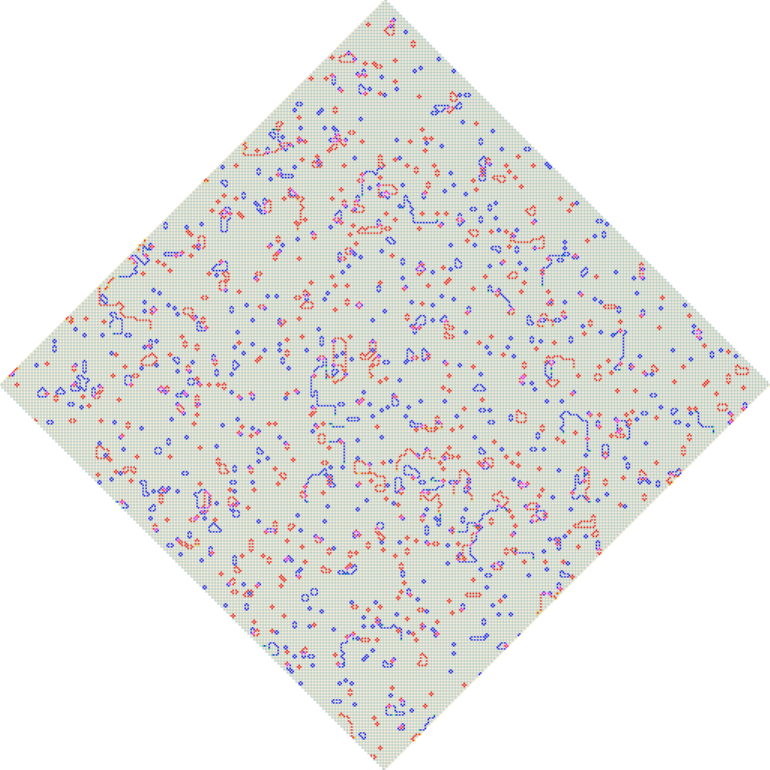}}
\end{center}
\caption{
    The distance 257 rotated surface code subject to depolarising noise (from Figure~\ref{fig:rot_12_flowchart3}) after five passes through the ANN decoder.
}\label{fig:rot_12_flowchart3}
\end{figure*}

\begin{figure*}[htp]
\begin{center}
\resizebox{170mm}{!}{\includegraphics{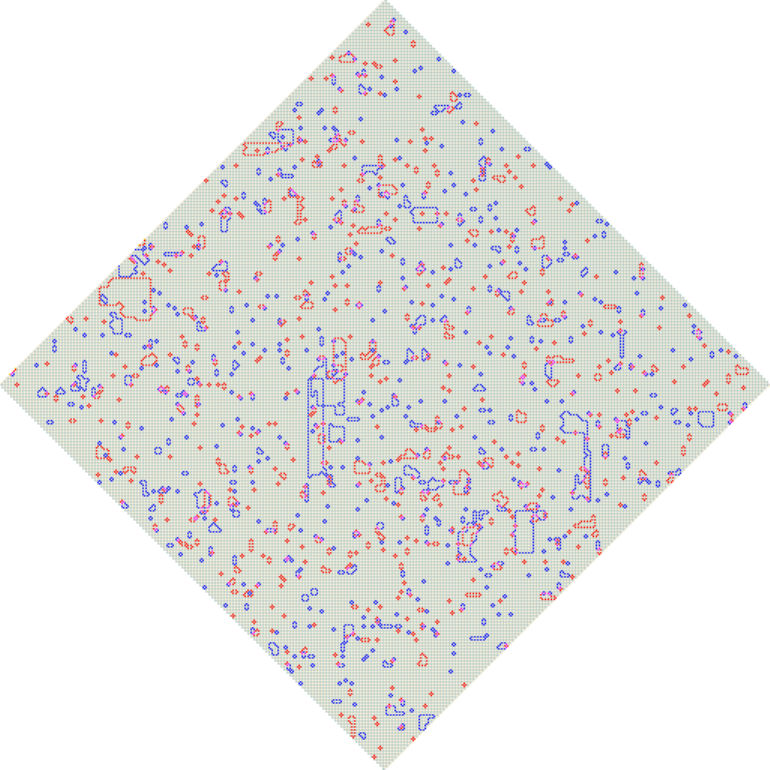}}
\end{center}
\caption{
    The distance 257 rotated surface code subject to depolarising noise (from Figure~\ref{fig:rot_12_flowchart1}) after five passes through the ANN decoder, followed by mop-up by the HDRG decoder.
}\label{fig:rot_12_flowchart4}
\end{figure*}

\begin{figure*}[htp]
\begin{center}
\resizebox{170mm}{!}{\includegraphics{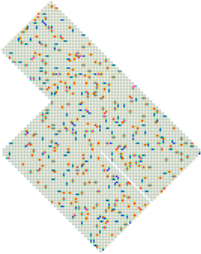}}
\end{center}
\caption{
    An example of a distance 33 surface code joint measurement lattice surgery structure subject to randomly distributed depolarising noise (8\%).
}\label{fig:mXXX_1}
\end{figure*}

\begin{figure*}[htp]
\begin{center}
\resizebox{170mm}{!}{\includegraphics{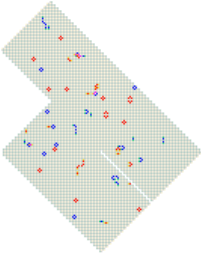}}
\end{center}
\caption{
    The distance 33 surface code joint measurement structure (from Figure~\ref{fig:mXXX_1}) after a single pass through the ANN decoder. 
}\label{fig:mXXX_2}
\end{figure*}

\end{document}